\documentclass[12pt,aps,prd,tightenlines,groupedaddress,preprint,floatfix,nofootinbib]{revtex4-1}

\usepackage{graphicx}
\usepackage{amssymb}
\usepackage{epstopdf}
\usepackage{feynmf}
\usepackage{hyperref}

\newcommand{\PRE}[1]{{#1}} 

\def\be{\begin{equation}}
\def\ee{\end{equation}}
\def\bea{\begin{eqnarray}}
\def\eea{\end{eqnarray}}

\def\tev{\, {\rm TeV}}
\def\gev{\, {\rm GeV}}
\def\mev{\, {\rm MeV}}
\def\kev{\, {\rm keV}}

\newcommand{\gsim}{\lower.7ex\hbox{$\;\stackrel{\textstyle>}{\sim}\;$}}
\newcommand{\lsim}{\lower.7ex\hbox{$\;\stackrel{\textstyle<}{\sim}\;$}}

\begin{document}

\setlength{\baselineskip}{0.25in}

\setcounter{footnote}{0}
\setcounter{page}{1}
\setcounter{figure}{0}
\setcounter{table}{0}


\preprint{UH-511-1252-2015}
\preprint{SCIPP-15/14}

\title{\PRE{\vspace*{0.8in}}
Constraints on Light Dark Matter from Single-Photon Decays of Heavy Quarkonium 
\PRE{\vspace*{0.3in}}
}
\author{Nicolas Fernandez}
\affiliation{Department of Physics and Santa Cruz Institute for Particle Physics, University of California, Santa Cruz, CA 95064 USA
\PRE{\vspace*{.5in}}
}

\author{Ilsoo Seong}
\affiliation{Department of Physics and Astronomy, University of Hawaii, Honolulu, HI 96822 USA
\PRE{\vspace*{.5in}}
}

\author{Patrick Stengel}
\affiliation{Department of Physics and Astronomy, University of Hawaii, Honolulu, HI 96822 USA
\PRE{\vspace*{.5in}}
}

\begin{abstract}
\PRE{\vspace*{.3in}}
We investigate constraints on the interactions of light dark matter with Standard Model quarks in a framework with effective contact operators mediating the decay of heavy flavor bound state quarkonium to dark matter and a photon. When considered in combination with decays to purely invisible final states, constraints from heavy quarkonium decays at high intensity electron-positron colliders can complement missing energy searches at high energy colliders and provide sensitivity to dark matter masses difficult to probe at direct and indirect detection experiments. We calculate the approximate limits on the branching fraction for $\Upsilon (1 S)$ decays to dark matter and a photon. Given the approximate limits on the branching fractions for all dimension 6 or lower contact operators, we present the corresponding limits on the interaction strength for each operator and the inferred limits on dark matter-nucleon scattering. Complementary constraints on dark matter annihilation from gamma-ray searches from dwarf spheroidal galaxies are also considered.         
\end{abstract}


\maketitle

\section{Introduction}

One of the strongest motivations for physics beyond the Standard Model (SM) is the observational evidence for gravitationally interacting non-baryonic dark matter (DM)~\cite{Kolb:1990}. If dark matter is permitted interactions with the SM, as predicted by many SM extensions, then possible signatures of these interactions could arise in direct, indirect and collider dark matter searches. Furthermore, if the scale of the new physics mediating the dark matter-Standard Model interactions is large relative to the energies relevant to the various dark matter searches, then this ultraviolet (UV) scale can be integrated out of the full Lagrangian theory and we can relate potential dark matter signatures using the contact operator approximation in a generalized effective field theory (EFT)~\cite{Goodman:2011,Hooper:2009,Cao:2011,Kumar:2013iva,Rajaraman:2013,Dreiner:2013,Buckley:2013,Rajaraman:2013-2,
Busoni:2014,Buchmueller:2014,Busoni:2014-2,
Alves:2014,Fedderke:2014,D'Eramo:2014aba}.   

A well motivated class of dark matter candidates comes from models with light dark matter (LDM). As opposed to the more ubiquitous weakly interacting massive particles (WIMPs), LDM particles have masses typically below the weak scale and, thus, could produce nuclear recoils at direct detection experiments with energies near or below threshold for detection. Assuming the contact operator approximation is valid, we can use EFT to relate complementary bounds on dark matter-Standard Model interactions from monojet searches at high energy colliders~\cite{Feng:2005gj,Goodman:2010,Birkedal:2004,Hooper:2010,Bai:2010,Rajaraman:2011,Fox:2012,Bai:2013,Agrawal:2013,
Goodman:2011-2,Bai:2011,Papucci:2014,Kumar:2015wya} to limits on LDM scattering off of nuclei, possibly constraining dark matter-nucleon interactions for dark matter masses beyond the reach of direct detection. In a similar fashion, searches for invisible bound state decays at high luminosity colliders can constrain LDM interactions with the SM independent of any particular UV physics model. 

Previous studies have considered a variety of invisible bound state decays ~\cite{Essig:2013,Fayet:2007,Fayet:2010,Cotta:2013,Schmidt-Hoberg:2013hba,McElrath:2007,Badin:2010,McKeen:2009rm} and, in a related work, the authors of~\cite{Fernandez:2014eja} have thoroughly explored the complementary aspects of $\Upsilon (1S)$ and $J / \Psi$ decays to purely invisible final states. In this work, we explore constraints from the decay $\Upsilon (1S) \rightarrow \gamma + invisible$ and explore the possible complementarity of searches between different bound state decays, as well as the relationship between disparate dark matter detection strategies. In particular, bound state quarkonium decays to $\gamma \bar X X$ are, at quark level in the matrix element, identical to monophoton searches. However, in the non-relativistic limit, many of the possible DM-SM interaction structures can be constrained by a particular combination of bound state meson decays to $\gamma + invisible$ final states and, thus, could possibly offer an important complement to monophoton searches. For simplicity and due to the relative lack of data available for other choices of initial state, we exclusively calculate constraints from the decays of $\Upsilon (1S)$. In principle, we will see that constraints from different combinations of  initial bound states and decay channels (i.e., $\rightarrow  invisible$ or $\rightarrow \gamma + invisible$) can constrain the same interaction structures. Results for $\Upsilon$ decays into final states with scalar dark matter were also presented in~\cite{Yeghiyan:2010}, although the constraints on dark matter-quark contact operators from radiative decays were approximated with limits from $\Upsilon (3S) \rightarrow \gamma + invisible$ decays assuming an on-shell mediator~\cite{Aubert:2008as}. 

In this paper, we use the limit on $\Upsilon (1S) \rightarrow \gamma \bar X X$ decays for a particular DM-SM interaction structure and dark matter spin to obtain limits on all contact operators, of dimension six or lower, coupling scalar, fermion or vector LDM to bottomonium. In section II, we discuss the current experimental constraints on bound state decays to $\gamma + invisible$ and on dark matter annihilation from observations of dwarf spheroidal galaxies, then calculate the relevant decay widths and annihilation cross sections, respectively. In section III, we present the associated limits on the strength of the DM-SM couplings and compare to results from monojet dark matter searches and direct dark matter detection.   

\section{Contact Operators and Constraints}

\begin{figure}[h]

\center\includegraphics[width=0.8 \textwidth]{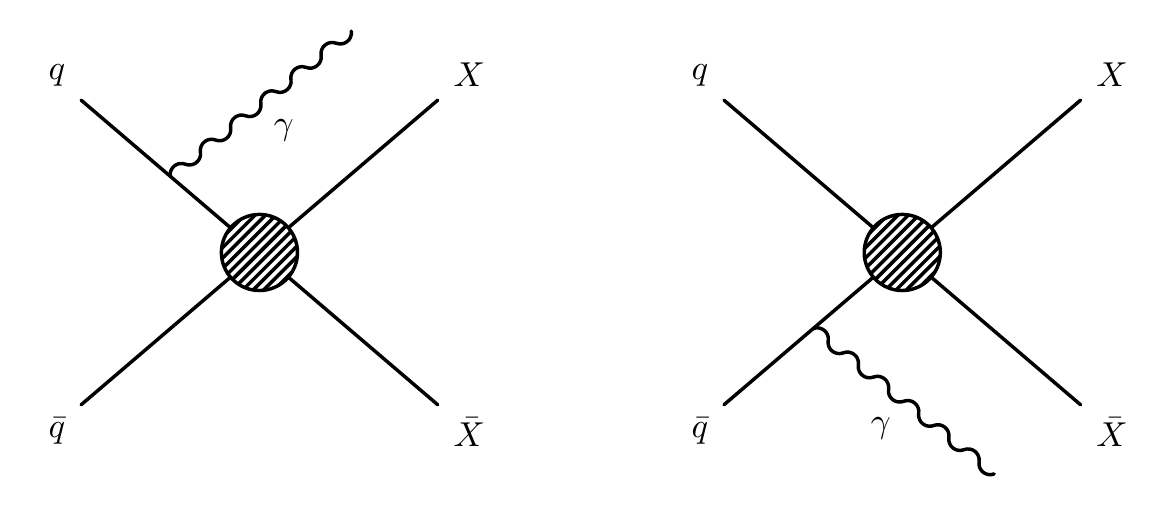}

\caption{Diagrams which yield the leading order contributions to matrix elements for quark/antiquark annihilation, ($\bar q q \rightarrow \gamma \bar X X$), and the corresponding matrix elements for bound state meson decays, ($\Upsilon (1S) \rightarrow \gamma \bar X X$).}

\label{fig:diagrams}

\end{figure}

We assume the dark matter-quark interaction is mediated by a UV physics model with heavy degrees of freedom that can be integrated out of the Lagrangian and, thus, can be parametrized as an effective four-point interaction, which can be written as Lorentz-invariant contractions of quark and dark matter bilinear structures. While the matrix element for quark/antiquark annihilation, ($\bar q q \rightarrow \gamma \bar X X$), and the corresponding matrix element for bound state meson decay can, in principle, receive contributions from amplitudes with a photon emitted by our heavy mediator, these contributions will necessarily be suppressed by extra powers of the UV scale relative to diagrams with photons emitted by the initial state quarks~\cite{Yeghiyan:2010}. We show the diagrams which yield the leading order contributions to the matrix elements for our bound state meson decays in Figure~\ref{fig:diagrams}.
     
In contrast to the contact operators considered in purely invisible decays, the quark bilinears in the operators mediating decays to dark matter with a photon in the final state need not share the angular momentum and $C/P$ properties of the heavy quarkonium state. Assuming weak interactions are negligible, the charge conjugation transformation required of the quark bilinears in the DM-SM interaction structures is determined by the presence or absence of a photon in the final state.
For the purely invisible decays of the $\Upsilon (1S)$ with $J^{PC} = 1^{--}$,  
 the quark bilinear must be $\bar q \gamma^i q$ or $\bar q \sigma^{0i} q$, where $i$ is a spatial index~\cite{Kumar:2013iva}.
  For $\Upsilon (1S) \rightarrow \gamma \bar X X $ decays, the final state photon is intrinsically $C$-odd and, thus, the associated quark bilinears in the operators mediating single-photon decays should be even under charge conjugation.
Therefore, the quark bilinear for heavy quarkonium decays to $\rightarrow \gamma + invisible$ must be $\bar q q $, $\imath \bar q \gamma^5 q $, $\bar q \gamma^0 \gamma^5 q$ or $\bar q \gamma^i \gamma^5 q$ and the effective operators which can yield a nonvanishing matrix element for $\Upsilon (1S) \rightarrow \gamma \bar X X$ decays are orthogonal to the operators which allow $\Upsilon (1S) \rightarrow \bar X X$ decays. When single-photon decays are considered in tandem with operators permitting purely invisible decays, all possible DM-SM interaction structures of dimension six or lower can be constrained by $\Upsilon$ decays. The interaction structures\footnote{In Tables~\ref{tab:structs} and~\ref{tab:structsINV}, we refer to spin-0, spin-1/2 and spin-1 dark matter fields with $\phi$, $X$ and $B^\mu$, respectively. In all other contexts, unless specifically noted, a dark matter field of arbitrary spin will be denoted $X$.} permitting $J^{PC} = 1^{--}$ bound state decays to $\gamma + invisible$ are listed in Table~\ref{tab:structs}, along with the angular momentum and $C/P$ properties of other possible bound states with non-vanishing matrix elements for decays to $\rightarrow  invisible$ or $\rightarrow \gamma + invisible$ final states\footnote{Note that, while the authors in~\cite{Fernandez:2014eja} had emphasized the complementarity between constraints on light dark matter from bound state decays to $invisible$ and other dark matter searches, here we note the relationship between the constraints arising from a variety of bound state decays. In the appendix, we update the constraints from purely invisible decays and calculate the relevant branching fractions and cross sections for the V7$_-$ and V9$_-$ operators not considered in previous work.\\ \, \\ \, \\}. If the interaction structures can permit $s$-wave dark matter annihilation, then a bound can also be set by Fermi observations of photons originating from dwarf spheroidal galaxies. Lastly, we indicate whether or not constraints on spin-independent (SI) or spin-dependent (SD) nucleon scattering can be inferred from BaBar or Fermi limits.

\begin{table}[h]
\centering
\begin{tabular}{|c|c|c|c|c|c|}
\hline
 Name & Interaction Structure & Invisible & Radiative & Annihilation  & Scattering \\
\hline
 F1  & $ (m_q / \Lambda^3) \bar X  X \bar q  q$ & $(0^{+ +} , 1 , 1)$ & $(1^{- -} , 1 , 0)$ & No & SI \\
     & & & $(1^{+ -} , 0 , 1)$ & &    \\
\hline
 F2  & $ (m_q / \Lambda^3) \imath \bar X  \gamma^5 X \bar q  q$ & $(0^{+ +} , 1 , 1)$ & $(1^{- -} , 1 , 0) $ & Yes & No \\
     & & & $ (1^{+ -} , 0 , 1) $ &  &  \\
\hline
 F3  & $ (m_q / \Lambda^3) \imath  \bar X  X \bar q \gamma^5 q$ & $(0^{- +} , 0 , 0)$ & $(1^{- -} , 1 , 0) $& No & No \\
 & & & $ (1^{+ -} , 0 , 1) $ &  &  \\
\hline
 F4  & $ (m_q / \Lambda^3)  \bar X \gamma^5 X \bar q \gamma^5 q$ & $(0^{- +} , 0 , 0)$ & $(1^{- -} , 1 , 0) $ & Yes & No \\
 & & & $ (1^{+ -} , 0 , 1) $ &  &  \\
\hline
 F7  & $ (1/\Lambda^2) \bar X \gamma^\mu X \bar q \gamma_\mu \gamma^5 q$ & $(0^{- +} , 0 , 0)$ & $(1^{- -} , 1 , 0)$ & Yes & No \\
  & &  $ (1^{+ +} , 1 , 1) $ & &  &  \\
\hline
 F8  & $ (1/\Lambda^2) \bar X \gamma^\mu \gamma^5 X \bar q \gamma_\mu \gamma^5 q$ & $(0^{- +} , 0 , 0)$ & $(1^{- -} , 1 , 0)$ & Yes & SD \\
  & & $ (1^{+ +} , 1 , 1) $ & &  &  \\
\hline
 S1  & $ (m_q / \Lambda^2)  \phi^{\dagger} \phi \bar q  q$ & $(0^{+ +} , 1 , 1)$ & $(1^{- -} , 1 , 0) $& Yes & SI \\
 & & & $ (1^{+ -} , 0 , 1) $ &  &  \\
\hline
 S2  & $ (m_q / \Lambda^2) \imath \phi^{\dagger} \phi \bar q \gamma^5 q$ & $(0^{- +} , 0 , 0)$ & $(1^{- -} , 1 , 0)$ & Yes & No \\
 & & & $ (1^{+ -} , 0, 1) $ &  &  \\
\hline
 S4  & $ (1/ 2 \Lambda^2) \imath ( \phi^{\dagger} \partial_\mu \phi -  \phi \partial_\mu  \phi^{\dagger} ) \bar q \gamma^\mu \gamma^5 q $ & $(0^{- +} , 0 , 0)$  &  $(1^{- -} , 1 , 0)$ & No & No \\ 
 & &  $ (1^{+ +} , 1 , 1) $ & &  &  \\
\hline
 V1  & $ (m_q / \Lambda^2)  B^{\dagger}_\mu B^\mu \bar q q$ & $(0^{+ +} , 1 , 1)$  & $(1^{- -} , 1 , 0) $ & Yes & SI \\
 & & & $ (1^{+ -} , 0 , 1) $ &  &  \\
\hline
 V2  & $ (m_q / \Lambda^2) \imath B^{\dagger}_\mu B^\mu \bar q \gamma^5 q$ & $(0^{- +} , 0 , 0)$ & $(1^{- -} , 1 , 0) $ & Yes & No \\
 & & & $ (1^{+ -} , 0 , 1) $ &  &  \\
\hline
 V4  & $ (1/ 2 \Lambda^2) \imath ( B_\nu^{\dagger} \partial_\mu B^\nu -  B^\nu \partial_\mu B_\nu^{\dagger} ) \bar q \gamma^\mu \gamma^5 q$ & $(0^{- +} , 0 , 0)$ &  $(1^{- -} , 1 , 0)$ & No & No \\
 & &  $ (1^{+ +} , 1 , 1) $ & &  &  \\
 \hline
 V8$_+$  & $ (1/ 2 \Lambda^2)( B_\nu^{\dagger} \partial^\nu B_\mu +  B_\nu \partial^\nu B_\mu^{\dagger}  ) \bar q \gamma^\mu \gamma^5 q$ & $(0^{- +} , 0 , 0)$ & $(1^{- -} , 1 , 0)$ & Yes & No \\
 & &  $ (1^{+ +} , 1 , 1) $ & &  &  \\
\hline
 V8$_-$  & $ (1/ 2 \Lambda^2) \imath ( B_\nu^{\dagger} \partial^\nu B_\mu -  B_\nu \partial^\nu B_\mu^{\dagger}  ) \bar q \gamma^\mu \gamma^5 q$ & $(0^{- +} , 0 , 0)$ & $(1^{- -} , 1 , 0)$ & No & No \\
 & &  $ (1^{+ +} , 1 , 1) $ & &  &  \\
\hline
 V10$_+$   & $ (1/ 2 \Lambda^2) \epsilon^{ \mu \nu \rho \sigma} ( B_\nu^{\dagger} \partial_\rho B_\sigma +  B_\nu \partial_\rho B_\sigma^{\dagger}  ) \bar q \gamma_\mu \gamma^5 q$ & $(0^{- +} , 0 , 0)$ &  $(1^{- -} , 1 , 0)$ & No & SD \\
 & &  $ (1^{+ +} , 1 , 1) $ & &  &  \\
 \hline
 V10$_-$   & $ (1/ 2 \Lambda^2) \imath \epsilon^{ \mu \nu \rho \sigma} ( B_\nu^{\dagger} \partial_\rho B_\sigma -  B_\nu \partial_\rho B_\sigma^{\dagger}  ) \bar q \gamma_\mu \gamma^5 q$ & $(0^{- +} , 0 , 0)$ &  $(1^{- -} , 1 , 0)$ & Yes & No \\
 & &  $ (1^{+ +} , 1 , 1) $ & &  &  \\
\hline
\end{tabular}
\caption{Interaction structures that can mediate $\Upsilon (1S)$ decays to $invisible + \gamma$. Invisible (radiative) identifies the $(J^{PC}, S, L)$ bound states that can be annihilated for $invisible$ ($invisible + \gamma$) final states. Note that we only consider $s$- and $p$-wave bound states. If the interaction structures can permit $s$-wave dark matter annihilation, then a bound can also be set by Fermi observations of photons originating from dwarf spheroidal galaxies. Lastly, we indicate whether or not constraints on spin-independent (SI) or spin-dependent (SD) nucleon scattering can be inferred from BaBar or Fermi limits. }
\label{tab:structs}
\end{table}

While the authors in~\cite{Fernandez:2014eja} had considered the complementarity between heavy quarkonium decays and other dark matter detection strategies, Table~\ref{tab:structs} (and Table~\ref{tab:structsINV}, located in the appendix) makes clear that the various DM-SM interaction structures will, in principle, have complementarity constraints due to decays of different bound states to invisible final states. For instance, assuming fermionic dark matter interacts through our F4 operator, which arises from integrating out a pseudoscalar mediator, constraints from $\Upsilon (1S) \rightarrow \gamma \bar X X$ decays will be directly related $h_b (1P) \rightarrow \gamma \bar X X$ and $\eta_{b} (1S) \rightarrow \bar X X$ decays. While calculation of decays beyond $\Upsilon (1S)$ is beyond the scope of this paper and constraints on invisible decays for less ubiquitous bound states may pose experimental challenges, we note the relationships between different bound state decays make high luminosity colliders an invaluable tool in determining the nature of SM interactions with LDM.          

\subsection{$\Upsilon (1S)$ Decays}

Since we only consider the decay of $s$-wave meson bound states in the nonrelativistic limit, the width for decays to any particular final state should be proportional to the value of the bound state wave function squared at the origin, $ | \psi (0) |^2 $. For the $\Upsilon (1S)$, the value of the bound state wave function squared at the origin can then be determined from the branching fraction to $e^+ e^-$,
\bea
{\cal B} ( \Upsilon (1S) \rightarrow e^{+} e^{-} ) = 16 \pi\alpha^2 Q_b^2 { | \psi_\Upsilon (0) |^2 \over \Gamma_\Upsilon M^2 }
= 0.0238 \pm 0.0011 ,
\eea
assuming the photon exchange contribution to the decay dominates the contribution from $Z$, $h$-exchange, with $M = 9460.30 \pm 0.26 \mev$, $\Gamma_\Upsilon = 54.02 \pm 1.25 \kev$~\cite{PDG}. We can also calculate the SM contribution for $\Upsilon (1S)$ decays to $\gamma + invisible$~\cite{Yeghiyan:2010}, yielding   
\bea
{\cal B} ( \Upsilon (1S) \rightarrow \bar \nu \nu \gamma ) = 2.48 \times 10^{-9} .
\eea
As we will show, this contribution is small relative to the current limits set by CLEO~\cite{Balest:1994ch} and BaBar~\cite{delAmoSanchez:2010ac} on $\Upsilon (1S) \rightarrow \gamma \bar X X$ for any DM-SM interaction structure and, thus, can be ignored. These searches operate at the $\Upsilon (2S)$ resonance and use the transition $\Upsilon (2S) \rightarrow \pi^+ \pi^- \Upsilon (1S)$ to identify $\Upsilon (2S)$ decays and reconstruct the $\Upsilon (1S)$ peak in the recoil mass distribution, $M_{rec}$, by tagging two oppositely charged pions with kinematics
\bea
M_{rec}^{2} \equiv s + M_{\pi \pi }^{2}-2\sqrt{s}E^{*}_{\pi \pi },
\eea  
where $E^*_{\pi \pi}$ is the energy of the dipion system in the center-of-mass (CM) frame of the $\Upsilon(2S)$, $M_{\pi \pi}$ is the invariant mass of the dipion system and $\sqrt{s} = 10023.26 \pm 0.31 \mev$ is the $\Upsilon(2S)$ resonance energy~\cite{PDG}. In addition to the pair of charged pion tracks, event selection, in the most recent analysis~\cite{delAmoSanchez:2010ac}, requires a single energetic photon with $E^*_{\gamma} \ge 150 \mev$ and $-0.73 < \cos \theta_\gamma^* < 0.68$, in the CM frame of the $\Upsilon(2S)$.  The limits on $\Upsilon (1S) \rightarrow \gamma \bar X X$ decays also assume a DM-SM interaction structure with a coupling between quarks and spin-0 dark matter which could be modeled by either our S1 or S2 interaction structure. The S1 and S2 interaction structures, we shall see, yield identical decay widths.  

We can calculate the branching fractions for $\Upsilon (1S) \rightarrow \gamma \bar X X$ given the relevant DM-SM interaction structures.      
We present the fully integrated branching fractions as functions of dark matter mass, $\Upsilon (1S)$ mass, mediation scale, the branching fraction to $e^+ / e^-$, and phase space integrals over the invariant mass squared of the invisible system, which are expressed analytically in the appendix. Note that, for the fully integrated branching fractions, the invariant mass squared of the invisible system, $X^2$, should be integrated over the entire kinematically allowed interval, $4 m_X^2 \le X^2 \le M^2$. The relevant branching fractions are
\bea
{\cal B}_{F1,F3} ( \gamma \bar X X)
&=&  {{\cal B} ( e^+ e^- ) M^2 m_X^4 \over 4 \pi^3 \alpha \Lambda^6}
\left[  I_{3/2}^{1}
- 4 { m_X^2 \over M^2} I_{3/2}^{2}
 \right] ,
 \nonumber\\
{\cal B}_{F2,F4} ( \gamma \bar X X)
&=&  {{\cal B} ( e^+ e^- ) M^2 m_X^4 \over 4 \pi^3 \alpha \Lambda^6}
\left[  I_{1/2}^{1}
- 4 { m_X^2 \over M^2} I_{1/2}^{2}
 \right] ,
 \nonumber\\
{\cal B}_{F7} ( \gamma \bar X X)
&=&  {{\cal B} ( e^+ e^- ) M^2 m_X^2 \over 12 \pi^3 \alpha \Lambda^4}
 \left[
 - 4 {m_X^2 \over M^2} I_{3/2}^{1} +  16 {m_X^4 \over M^4} I_{3/2}^{2} + I_{1/2}^{-1} 
\right.
\nonumber\\
&\,& ~~~~~~~~~~~~~~~~~~~~~~ \left.  
 +  \left( 2 - 4 {m_X^2 \over M^2}  \right) I_{1/2}^0
+ 4 {m_X^2 \over M^2} I_{1/2}^{1} - 48 { m_X^4 \over M^4} I_{1/2}^{2}
 \right] ,
  \nonumber\\
{\cal B}_{F8} ( \gamma \bar X X)
&=&  {{\cal B} ( e^+ e^- ) M^2 m_X^2 \over 12 \pi^3 \alpha \Lambda^4}
 \left[
 8 { m_X^2 \over M^2} I_{3/2}^{1} - 32 { m_X^4 \over M^4} I_{3/2}^{2} +
  I_{1/2}^{-1} +  \left( 2 - 4 {m_X^2 \over M^2} \right) I_{1/2}^{0}
- 8 {m_X^2 \over M^2} I_{1/2}^{1}
 \right] ,
 \nonumber\\
{\cal B}_{S1,S2} ( \gamma \bar X X)
&=&  {{\cal B} ( e^+ e^- ) M^2 m_X^2 \over 32 \pi^3 \alpha \Lambda^4}
\left[ I_{1/2}^{0} - 4 {m_X^2 \over M^2} I_{1/2}^{1} \right] ,
\nonumber\\
{\cal B}_{S4} ( \gamma \bar X X)
&=&  {{\cal B} ( e^+ e^- ) M^2 m_X^2 \over 96 \pi^3 \alpha \Lambda^4}
   \left[ I_{3/2}^{0} - 16 {m_X^4 \over M^4} I_{3/2}^{2} \right]  ,
\nonumber\\
{\cal B}_{V1,V2} ( \gamma \bar X X)
&=&  {{\cal B} ( e^+ e^- ) M^2 m_X^2 \over 32 \pi^3 \alpha \Lambda^4}
\left[ 
3 I_{1/2}^{0} 
+ \left(- 4  - 12 {m_X^2 \over M^2} \right) I_{1/2}^{1} 
+ \left(4  + 16 {m_X^2 \over M^2} \right) I_{1/2}^{2} 
- 16 {m_X^2 \over M^2}  I_{1/2}^{3}
\right] ,
\nonumber\\
{\cal B}_{V4} ( \gamma \bar X X)
&=&  {{\cal B} ( e^+ e^- ) M^2 m_X^2 \over 96 \pi^3 \alpha \Lambda^4}
\left[ 
3 I_{3/2}^{0} 
- 4   I_{3/2}^{1} 
+ \left( 4  - 48 {m_X^4 \over M^4} \right) I_{3/2}^{2} 
+  64 {m_X^4 \over M^4} I_{3/2}^{3} 
- 64 {m_X^4 \over M^4}  I_{3/2}^{4}
\right] ,
\nonumber\\
{\cal B}_{V8_+} ( \gamma \bar X X)
&=&   {{\cal B} ( e^+ e^- ) M^2 m_X^2 \over 24 \pi^3 \alpha \Lambda^4}
\left[ 
- 2 I_{3/2}^1 + \left( 12 {m_X^2 \over M^2} + 3  \right) I_{3/2}^2 - \left( 16 {m_X^4 \over M^4 } + 12 {m_X^2 \over M^2}  \right) I_{3/2}^3
  \right],
\nonumber\\
{\cal B}_{V8_-} ( \gamma \bar X X)
&=& 
 {{\cal B} ( e^+ e^- ) M^2 m_X^2 \over 24 \pi^3 \alpha \Lambda^4}
\left[ 
 I_{3/2}^{1} +    I_{3/2}^{2}  - 16 {m_X^4 \over M^4} I_{3/2}^{3} - 16 {m_X^4 \over M^4} I_{3/2}^{4}
\right] ,
\nonumber\\
{\cal B}_{V10_+} ( \gamma \bar X X)
&=&  {{\cal B} ( e^+ e^- ) M^2 m_X^2 \over 48 \pi^3 \alpha \Lambda^4}
\left[ 
 I_{3/2}^{0} + \left( 2 - 12  {m_X^2 \over M^2} \right) I_{3/2}^{1}  + 32 {m_X^4 \over M^4} I_{3/2}^{2} - 32 {m_X^4 \over M^4} I_{3/2}^{3}
\right] ,
\nonumber\\
{\cal B}_{V10_-} ( \gamma \bar X X)
&=& 
 {{\cal B} ( e^+ e^- ) M^2 m_X^2 \over 48 \pi^3 \alpha \Lambda^4}
\left[ 
 I_{1/2}^{0} +  2  I_{1/2}^{1}  - 16 {m_X^4 \over M^4} I_{1/2}^{2} - 32 {m_X^4 \over M^4} I_{1/2}^{3}
\right] ,
\label{eqn:fullwidths}
\eea
with phase space integrals, $I^{m}_{n}$, defined in Appendix \ref{phase_space}. Note that, for compactness, we have written the branching fractions in a form which somewhat obscures the $m_X$ dependence and that the phase space integrals are defined such that one can approximate the scaling $I^{m}_{n} \propto m_X^{-2(m+1)}$ in the low mass limit. We have approximated throughout the calculation of the branching fractions that $M \simeq 2 m_q$, ignoring the ${\cal O} ( \Lambda_{QCD}) $ mass difference~\cite{Yeghiyan:2010}.  For operators F1, F2, F3, F4 and F8, we assume the dark matter is a Dirac fermion. Alternatively, if the dark matter were a Majorana fermion, the branching fractions will be larger by a factor of $2$ and the F7 operator would vanish. Similarly, we have assumed complex fields for operators mediating quark interactions with spin-0 and spin-1 dark matter. Conversely, if we assumed real dark matter fields, operators S4, V4, V8$_-$ and V10$_-$ would vanish and the remaining operators will have a larger branching fraction by a factor of $2$.  The branching fractions for spin-1 dark matter have terms proportional to $m_X^{-2}$ or $m_X^{-4}$ due to the longitudinal polarization modes of the dark matter. As discussed in~\cite{Fernandez:2014eja}, constraints due to the unitarity of the associated matrix elements are trivial for the decay of bound state mesons, which are approximated to be non-relativistic, compared to the application of unitarity constraints to monojet searches for spin-1 dark matter at LHC~\cite{Kumar:2015wya}. Note that the branching fractions for S1, S2 and S4 match those previously calculated in~\cite{Yeghiyan:2010}.         

\begin{figure}[h]
\includegraphics[width=\textwidth]{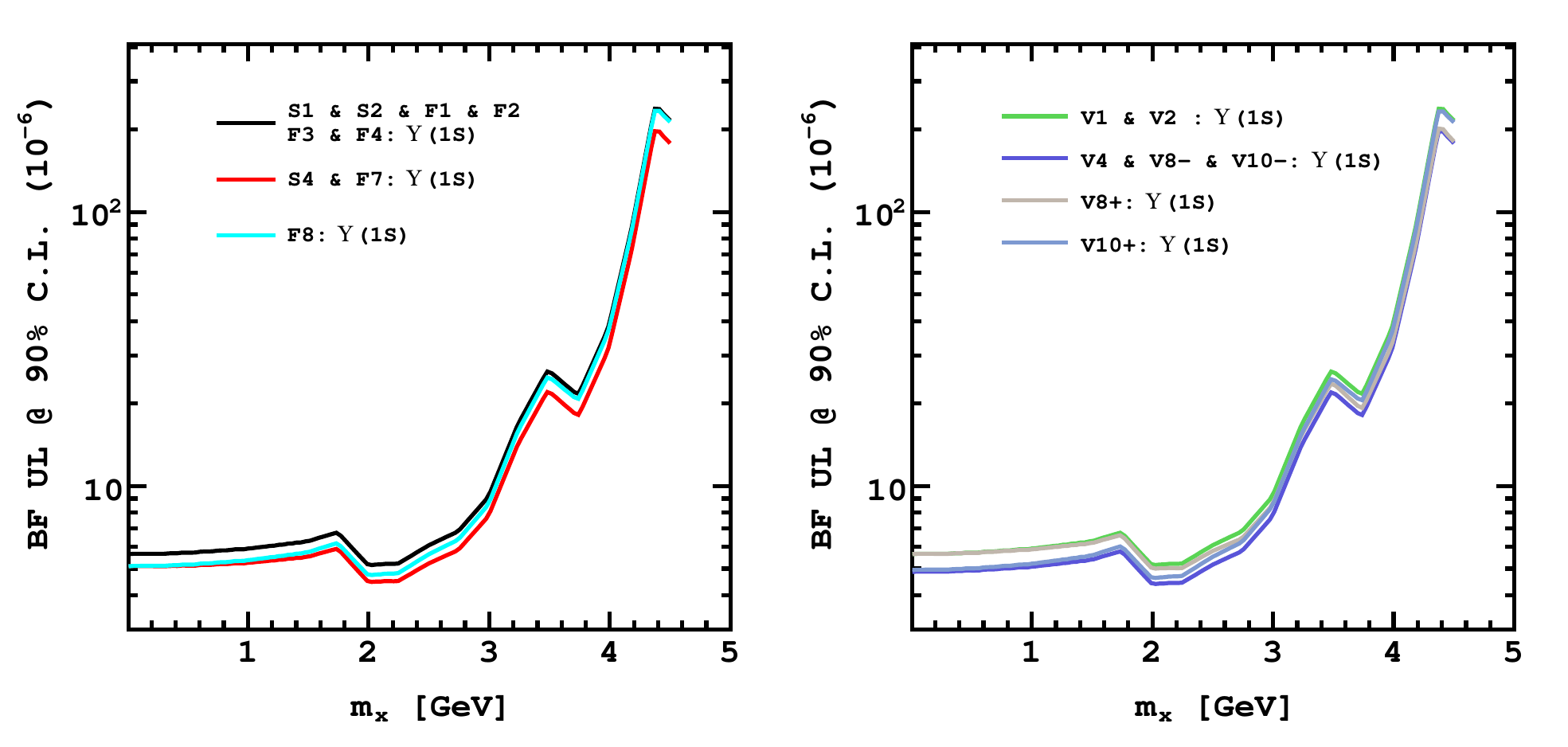}
\caption{Approximate limits on ${\cal B} (\Upsilon (1S) \rightarrow \gamma \bar X X )$ for all relevant effective contact operators mediating interactions with spin-$0$ (left), spin-$1/2$ (left) and spin-$1$ (right) dark matter rescaled from 90\% CL BaBar limits~\cite{delAmoSanchez:2010ac}. }
\label{fig:BF_limit}
\end{figure}

We also calculate the differential branching fractions for $\Upsilon (1S) \rightarrow \gamma \bar X X$, with respect to photon energy and scattering angle, assuming polarized $\Upsilon (1S)$ produced in $\Upsilon (2S) \rightarrow \pi^+ \pi^- \Upsilon (1S)$ transitions. For relativistic $e^+ e^-$ annihilating through a photon, the resulting $\Upsilon (2S)$ is produced polarized with its spin axis lying along the beam line. The daughter $\Upsilon (1S)$ should also be polarized along the beam line and we assume a negligible differential boost between the rest frame of the $\Upsilon (2S)$ and that of the $\Upsilon (1S)$.
If we partially integrate the polarized differential branching fractions over the phase space of the photon given the relevant analysis cuts, denoted $\theta_0$ for scattering angle and $\omega_0$ for energy in the CM frame, we can approximate limits on the full branching fraction of any interaction structure given the constraints on our S1/S2 operators presented in~\cite{delAmoSanchez:2010ac}.  
We present the differential branching fractions for decays of polarized bound state mesons in the appendix and we denote the partially integrated branching fractions ${\cal B}_i^{pol} (\theta_0, \omega_0 )$,  where $i$ labels an interaction structure.
For any set of cuts on the photon phase space, we can define an efficiency for each interaction structure,  ${\cal F}_i ( \theta_0, \omega_0)  = {\cal B}_i^{pol} ( \theta_0, \omega_0) / {\cal B}_i $. We can multiply the limits on ${\cal B}_{S1,S2} (\gamma \bar X X)$~\cite{delAmoSanchez:2010ac} by ${\cal F}_{S1,S2}$, given the associated analysis cuts, and are left with approximate limits on a partial branching fraction which is proportional to the number of events observed in the detector, independent of effective operator. The limit on the full branching fraction for any operator is then given by the product of the limit on S1/S2 and the ratio ${\cal F}_{S1,S2} / {\cal F}_i$,  plotted in Figure~\ref{fig:BF_limit}.            

The rescaling of each operator is virtually insensitive to the photon energy threshold, $\omega_0$, and there are only small corrections for some operators due the geometric acceptance, $\theta_0$. As we demonstrate in the appendix, the angular phase space distributions for all scalar and pseudoscalar mediated interaction structures are identical and factorizable. Thus, in our approximation, the geometric acceptance of the S1/S2 branching fraction will exactly cancel that from the F1/F3, F2/F4 and V1/V2 branching fractions, yielding a nearly identical limit. Note that, our approximate rescaling of the S1/S2 limits assumes uniform detector efficiency within the phase space of the detector. While setting a true limit on each interaction structure would require taking these details into account, such analysis is beyond the scope of this work and our rescaled limits should be considered estimates of experimental sensitivity.        

\subsection{Dark Matter Annihilation}

We calculate the cross sections for ($\bar X X \rightarrow \bar q q$), given effective operators which have associated matrix elements that allow for $s$-wave dark matter annihilation. Note that only nine of the operators which allow $J^{PC}= 1^{--}$ bound state meson decays to $\gamma + invisible$ also allow for unsuppressed dark matter annihilation. The corresponding tree-level dark matter annihilation cross sections are

\bea
\langle \sigma_A^{F2} v \rangle &=& { 3 m_q^2 m_X^2 \over 2 \pi \Lambda^6 }  \left( 1 - { m_q^2 \over m_X^2} \right)^{3/2},
\nonumber\\
\langle \sigma_A^{F4} v \rangle &=& { 3 m_q^2 m_X^2   \over 2 \pi \Lambda^6 } \left( 1 - { m_q^2 \over m_X^2} \right)^{1/2},
\nonumber\\
\langle \sigma_A^{F7} v \rangle &=& { 3 m_X^2 \over  \pi \Lambda^4 }  \left( 1 - { m_q^2 \over m_X^2} \right)^{3/2} ,
\nonumber\\
\langle \sigma_A^{F8} v \rangle &=& { 3 m_q^2 \over 2  \pi \Lambda^4 }  \left( 1 - { m_q^2 \over m_X^2} \right)^{1/2},
\nonumber\\
\langle \sigma_A^{S1} v \rangle &=& { 3 m_q^2  \over 4 \pi \Lambda^4 } \left( 1 - { m_q^2 \over m_X^2} \right)^{3/2},
\nonumber\\
\langle \sigma_A^{S2} v \rangle &=& { 3 m_q^2  \over 4 \pi \Lambda^4 } \left( 1 - { m_q^2 \over m_X^2} \right)^{1/2},
\nonumber\\
\langle \sigma_A^{V1} v \rangle &=& { m_q^2  \over 4 \pi \Lambda^4 } \left( 1 - { m_q^2 \over m_X^2} \right)^{3/2},
\nonumber\\
\langle \sigma_A^{V2} v \rangle &=& { m_q^2  \over 4 \pi \Lambda^4 } \left( 1 - { m_q^2 \over m_X^2} \right)^{1/2},
\nonumber\\
\langle \sigma_A^{V10_-} v \rangle &=& { m_X^2 \over 3 \pi \Lambda^4 }  \left( 1 - { m_q^2 \over m_X^2} \right)^{3/2} .
\eea

Note that, while V8$_+$ allows for $s$-wave annihilation, the associated matrix element is suppressed by an additional factor of $v^2$ due to the time-like polarization of the spin-$1$ dark matter and will vanish in the non-relativistic limit~\cite{Kumar:2013iva}. If we assume a flavor structure which relates $b$-quark couplings to dark matter with light quark couplings, then 
$\Lambda$ can be constrained by a stacked analysis of the photon flux from dwarf spheroidal galaxies~\cite{Fermi:2011,Fermi:2013,Geringer:2011,Geringer:2012,Essig:2009,Ackermann:2015zua,Baring:2015sza}.
The photon flux is calculated as the product of a dark matter density factor, which can be inferred from the rotation curves of visible matter, and a particle physics flux, which can be expressed as
\bea
\Phi_{PP} &=& { \langle \sigma_A v \rangle \over 8 \pi m_X^2 } \int_{E_{thr}}^{m_X} \sum_f B_f {d N_f \over d E} d E .
\label{eqn:gammaflux}
\eea
For dark matter annihilation into a channel $f$ with branching ratio $B_f$ ,  $ d N_f / d E$ is the associated photon spectrum. Note that $E_{thr} = 1 \gev$ is
the energy threshold for the photon analysis. 

We produce the spectra for $\bar u u$, $\bar d d$ and $\bar s s$ annihilation channels in Pythia $6.403$~\cite{Pythia}, and then set bounds on $\Lambda$ given the $ 95 \% $ CL limit on $\Phi_{PP}$ from Fermi-LAT observations of dwarf spheroidal galaxies~\cite{Geringer:2011},
\bea
\Phi_{PP} < 5.0_{-4.5}^{+4.3} \times 10^{-30} \, {\rm cm}^3 \, {\rm s}^{-1} \gev^{-2}.
\eea
 Uncertainty in the dwarf halo profiles yields the asymmetric uncertainties, which are $95 \%$ CL systematic errors~\cite{Fermi:2011}. Note that, since we consider complex dark matter, we must weaken this bound by a factor $2$, as a real dark matter field was assumed in~\cite{Geringer:2011}.
Also note that, although more stringent limits on dark matter annihilation can be set by a more sophisticated analysis using more recent Fermi data~\cite{Ackermann:2015zua}, such an analysis is beyond the scope of this paper. We do however, note that, such analyses tend to improve limits on the dark matter annihilation cross section by factors of 2-10, depending on annihilation channel and sample of dwarf galaxies~\cite{Fermi:2011,Fermi:2013,Ackermann:2015zua}.  

\section{Results}

We assume scalar and pseudoscalar mediated interaction structures have effective couplings proportional to quark mass, while operators with pseudovector quark bilinears couple universally to all quark flavors. This particular choice of couplings is well motivated by any UV completion which assumes minimal flavor violation. For example, there is a class of models within the framework of the Next-to-Minimal Supersymmetric Standard Model (NMSSM) with light neutralino dark matter which will couple to quarks through a CP-odd Higgs boson~\cite{Gunion:2005rw}. We assume dark matter and quarks only interact through a single interaction structure, although this choice, as well as the flavor structure of the effective couplings, is only a benchmark and, in general, need not correspond to any particular UV completion. Also, while the effective couplings are not scale invariant, we assume the effect of RG-running from the scale of bound state decays to that of nuclear scattering to be negligible~\cite{Haisch:2013uaa,Crivellin:2014qxa}.         

\subsection{Mediator Scale}

\begin{figure}[h]
\includegraphics[width=\textwidth]{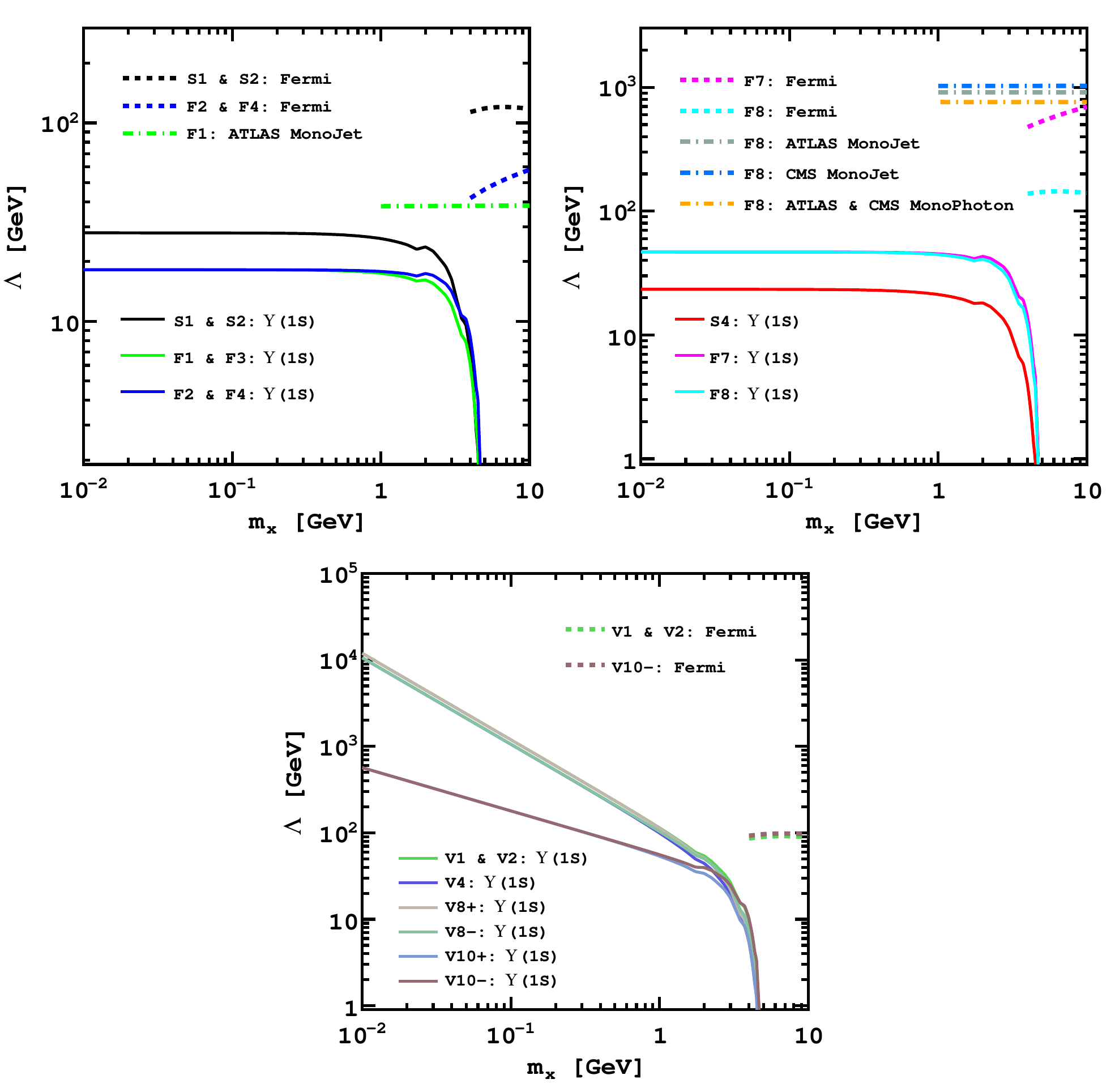}
\caption{Bounds on the mediator scale, $\Lambda$, for dark matter
of mass $m_X$ arising from constraints on $\Upsilon (1S) \rightarrow \gamma \bar X X$ decays, from constraints on dark matter annihilation
 in dwarf spheroidal galaxies, and from monojet/photon searches at ATLAS~\cite{Aad:2015zva,Aad:2014tda} and
CMS~\cite{Khachatryan:2014rra, Khachatryan:2014rwa}. The upper left panel shows bounds for scalar and pseudoscalar mediated interaction structures with scalar or fermionic dark matter.
The upper right panel shows constraints for pseudovector mediated interaction structures with scalar or fermionic dark matter.
The lower center panel shows limits for interaction structures with vector dark matter.  }
\label{fig:lambda}
\end{figure}

In Figure~\ref{fig:lambda} we plot bounds on $\Lambda$
arising from limits on invisible $\Upsilon (1S) \rightarrow \gamma \bar X X$ decays,
dark matter annihilation in dwarf spheroidal galaxies and monojet/photon searches at ATLAS~\cite{Aad:2015zva,Aad:2014tda} and
CMS~\cite{Khachatryan:2014rra, Khachatryan:2014rwa}.
The upper left panel shows bounds for scalar and pseudoscalar mediated interaction structures with scalar or fermionic dark matter.
The upper right panel shows constraints for pseudovector mediated interaction structures with scalar or fermionic dark matter.
The lower center panel shows limits for interaction structures with vector dark matter. 
Note that the bounds from LHC searches and dwarf spheroids primarily probe dark matter coupling to light flavor quarks, where constraints from $\Upsilon (1S) \rightarrow \gamma \bar X X$ decays directly probe couplings to $b$-quarks.    
Also, radiative bound state decays are analogous to monojet/photon searches at LHC, as the free quark annihilation matrix elements for $\bar q q \rightarrow \gamma \bar X X$ are identical, but evaluated at different energy scales. While the monojet constraints are only published for dark matter masses $m_X \gsim 1 \gev$, in principal these constraints are not threshold limited and extend to the massless dark matter limit with respective mass dependencies similar to those for the corresponding constraints from bound state decays.

Constraints on the effective mediator scale can be related to UV completions with heavy mediators of mass $M_{med} \sim  g \Lambda$, where $g$ is a dimensionless coupling constant for a given DM-SM interaction structure.
The effective contact approximation is only valid for mediators with masses larger than the momentum transfers which appear in the associated propagators.
 Thus, assuming $g \sim {\cal O} (1)$,
 the limits from bound state meson decays satisfy $M_{med} \gsim 10 \gev$, ensuring that the effective contact approximation is valid for all relevant contact operators.
Similarly, monojet searches at LHC can only set model independent bounds for mediator masses above $ \sim {\cal O} (\tev)$ and the range of validity for such constraints from bound state decays is complementary.
 Alternatively, if we consider simple UV completions of our effective operators outside of the mass range where our effective contact approximation is valid, monojet searches can set bounds on lighter mediator masses which are also beyond the sensitivity of constraints from bound state decays (for example, see~\cite{Haisch:2015ioa,Buchmueller:2014yoa}). However, we note that, even in simplified model frameworks, the sensitivity of monojet constraints to smaller mediator masses will eventually be limited by the large transverse energy required in such searches. For example, if we consider an $s$-channel UV completion, the typical momentum transfer, $\sqrt{\hat s}$, in monojet events should not be smaller than a few hundred $\gev$ at LHC8. 
In order to keep low $m_X$ monojet constraints on $\Lambda \sim {\cal O} (\tev)$ constant, 
 we must rescale $g$ to smaller values as we lower $M_{med}$.  
 For $M_{med}^2 \lsim \hat s $, the momentum transfer will dominate the propagator and the LHC monojet event rate will be suppressed by $\sim g^4 / \hat s^2 $.
In comparison, the constraints from bound state decays will stay constant with an event rate proportional to $\sim g^4 / M_{med}^4$. As a result, bound state decays in our effective field theory framework can constrain mediator masses in the range where LHC monojet limits, applied to simplified models, have limited sensitivity.       

The relative strength of the limits from bound state meson decays compared to those from LHC searches or dark matter annihilation is highly dependent on assumptions made about the effective couplings for our contact operators. As a benchmark, we have assumed scalar and pseudoscalar mediated interactions have couplings $\sim m_q$, which enhance the matrix elements for interactions with $b$-quarks relevant for $\Upsilon (1S) \rightarrow \gamma + invisible$ decays. For pseudovector mediated interaction structures constrained by $\Upsilon (1S) \rightarrow \gamma + invisible$, as well as the operators relevant for $\Upsilon (1S) \rightarrow invisible$ decays, we make no such assumption and the bounds are comparatively weak.
The associated suppression of the matrix elements for dark matter annihilation is manifest in the cross sections for F2, F4, S1, S2, V1, and V2 operators. 
 Also note that the limit on F8 from dwarf spheroids is somewhat weaker due to the chirality suppression of the dark matter annihilation matrix element.      
Astrophysical uncertainties can strengthen the constraints from dark matter annihilation in dwarf spheroidal galaxies by up to a factor of 10 or weaken them
by up to a factor of 2. 

\subsection{Dark Matter-Nucleon Scattering}

The effective operators which we consider that will yield velocity independent terms in
the respective dark matter-nucleon scattering matrix element are F1 (SI), F8 (SD), S1 (SI), V1 (SI) and V10$_+$ (SD).

For operators with scalar quark bilinears yielding spin-independent scattering, the associated dark matter-proton cross sections are
\bea
\sigma_{SI}^{F1} &=& {\mu_p^2 m_p^2  \over \pi \Lambda^6 } 
\left(\sum_{q=u,d,s} f_q^p + {2 \over 27} \sum_{c,b,t} f_g^p\right)^2 ,
\nonumber\\
\sigma_{SI}^{S1, V1} &=&  {\mu_p^2 m_p^2  \over 4 \pi \Lambda^4 m_X^2 } \left(\sum_{q=u,d,s} f_q^p + {2 \over 27} \sum_{c,b,t} f_g^p\right)^2 ,
\eea
where $m_p$ is the proton mass and $\mu_p$ is the reduced mass of the dark matter-nucleon system.
As a benchmark, we assume the nucleon form factors associated with the scalar quark bilinear are $f_u^p = f_d^n = 0.024 $, $ f_d^p = f_u^n = 0.035$, 
$ f_s^{p,n} = 0.051$ and $f_g^{p,n} = 1 - \sum_{q=u,d,s} f_q^{p,n} $~\cite{Kelso:2014qja}, although the precise determination of scalar nucleon couplings varies in the literature and there are possibly significant uncertainties due to the strangeness content of the nucleon (for recent discussions,~\cite{Alarcon:2011zs, Alarcon:2012nr, Hoferichter:2015dsa}). The relative strength of interactions between dark matter and different flavor quarks can have a significant impact on the overall scattering cross section. Here we have assumed operators with scalar quark bilinears have an effective coupling proportional to the quark mass, whereas in~\cite{Fernandez:2014eja}, the authors assume universal quark coupling for spin-independent scattering through interaction structures with vector quark bilinears. We also consider constraints on dark matter scattering through scalar or pseudoscalar mediated operators only assuming interactions with $b$-quarks.  

Similarly, for operators with pseudovector quark bilinears yielding spin-dependent scattering, the associated dark matter-proton cross sections are
\bea
\sigma_{SD}^{F8} &=& { 3 \mu_p^2 \over \pi \Lambda^4 }  
\left(\sum_{q=u,d,s}  \Delta_q^p \right)^2  ,
\nonumber\\
\sigma_{SD}^{V10_+} &=& { 2 \mu_p^2 \over  \pi \Lambda^4 }  
\left(\sum_{q=u,d,s}  \Delta_q^p \right)^2  .
\eea
The nucleon spin form factors associated with the pseudovector quark bilinear are
$\Delta_u^p = 0.84$, $\Delta_d^p = - 0.43$ and $\Delta_s^p = - 0.09$~\cite{Ellis:2009ai}. Note that, unlike spin-independent scattering, we only assume universal quark coupling to dark matter in the matrix elements yielding spin-dependent cross sections, as there is no significant coupling to heavy quark flavor in pseudovector mediated interactions.      

\begin{figure}[h]
\includegraphics*[width=0.49 \textwidth]{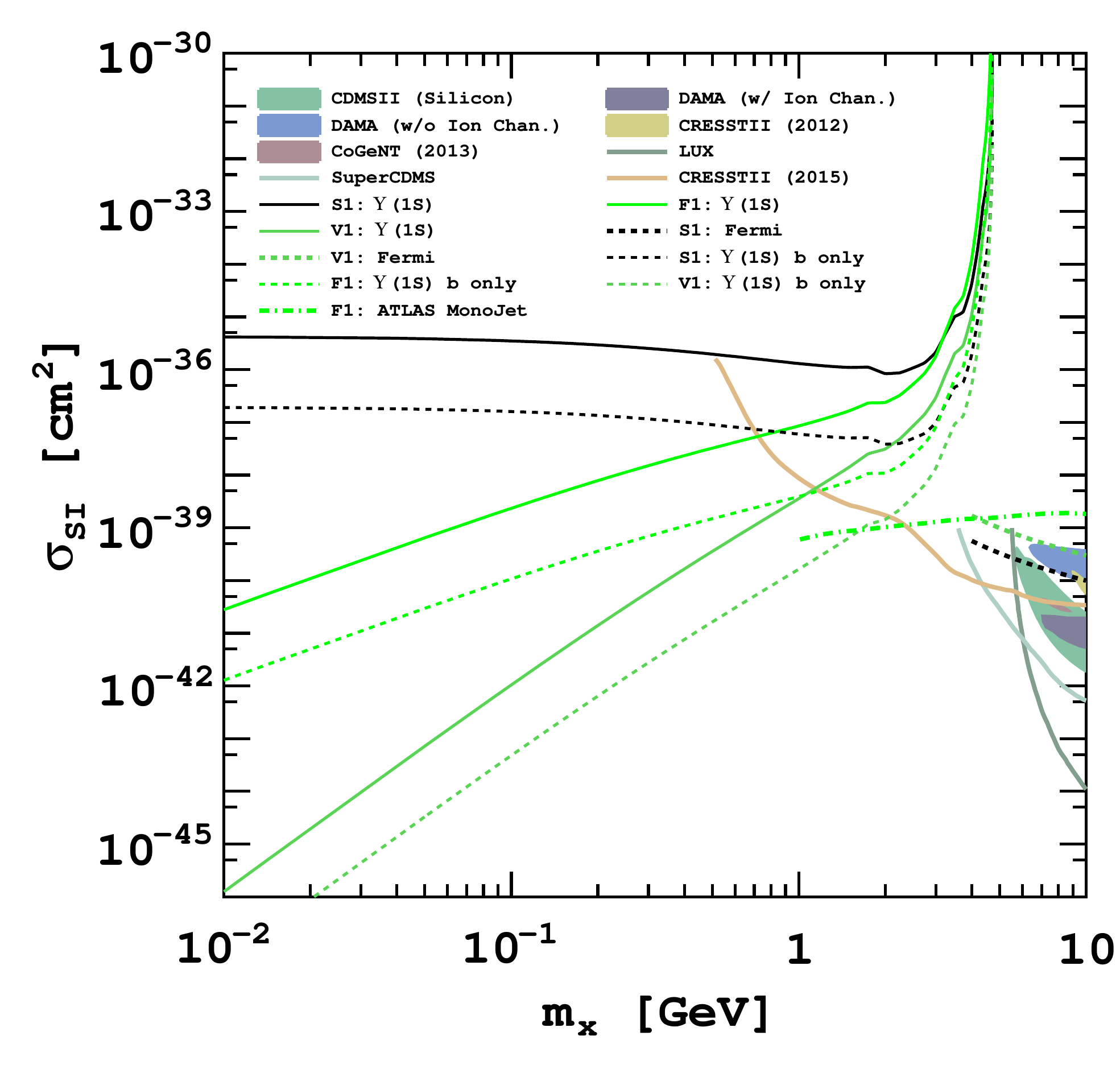}
\includegraphics*[width=0.49 \textwidth]{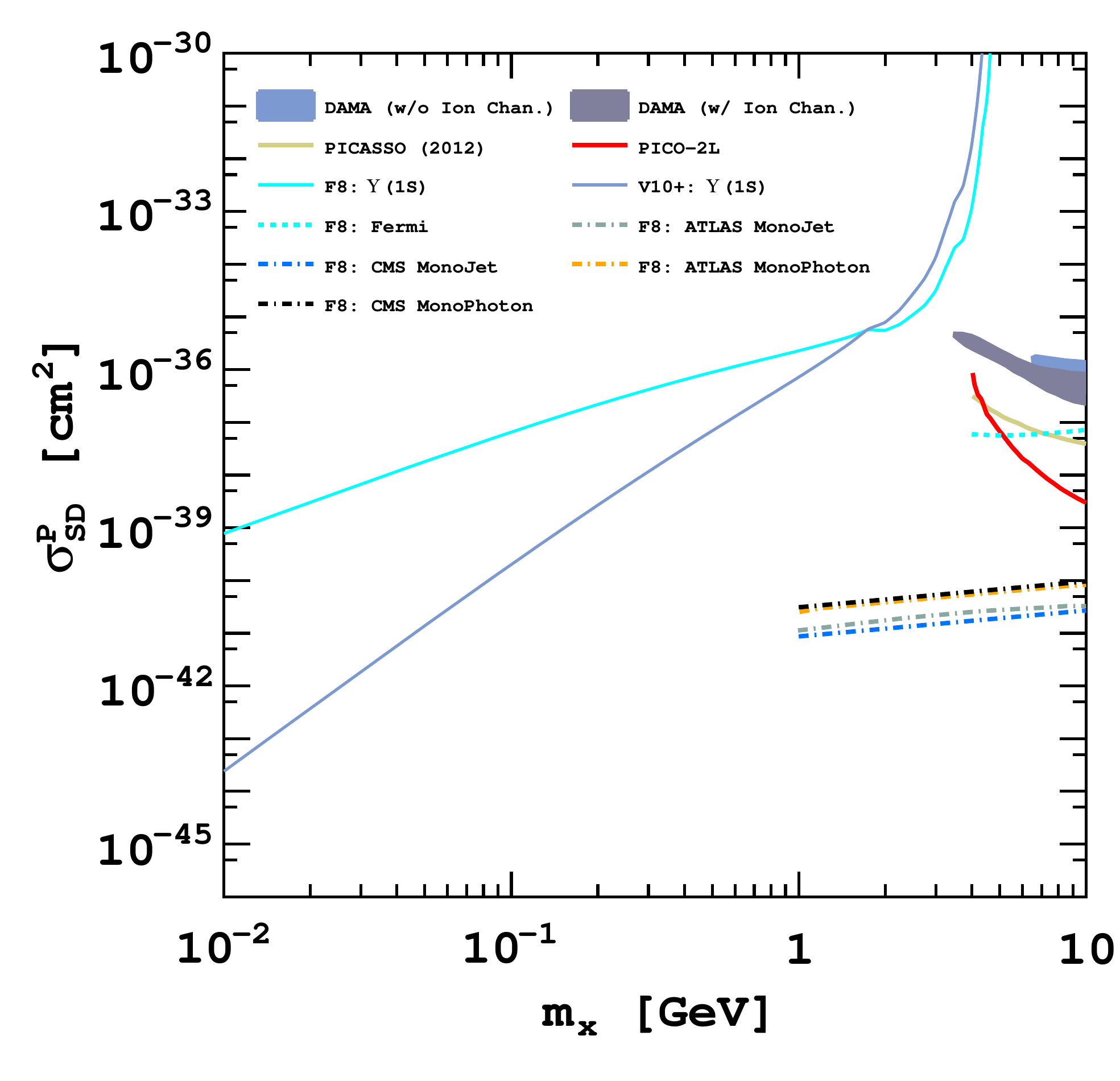}
\caption{
Bounds on the dark matter-proton spin-independent (left panel) and spin-dependent (right panel) scattering cross sections
for dark matter of mass $m_X$ coupling to quarks through the indicated effective contact operator.
 The solid (dashed) exclusion contours indicate 90\% CL bounds arising from limits on $\Upsilon (1S) \rightarrow \gamma \bar X X$ decays, assuming coupling to all (only $b$-flavor) quarks. The other labeled exclusion contours indicate
95\% CL bounds arising from Fermi-LAT constraints on dark matter annihilation in dwarf spheroidal galaxies,
and 90\% CL bounds arising from monojet and monophoton searches (ATLAS~\cite{Aad:2015zva,Aad:2014tda} and
CMS~\cite{Khachatryan:2014rra, Khachatryan:2014rwa}). The 90\% CL exclusion contours from CRESST II~\cite{Angloher:2015ewa},
SuperCDMS~\cite{CDMS:2014}, LUX~\cite{Akerib:2013tjd}, PICO~\cite{Amole:2015lsj} and PICASSO~\cite{Archambault:2012pm}
are also shown, as are the DAMA/LIBRA~\cite{Savage:2008er}, CRESST II (95\% CL)~\cite{Angloher:2011uu}, CoGeNT~\cite{Aalseth:2014jpa} and
CDMS II(Silicon)~\cite{Agnese:2013rvf} 90\% CL signal regions.}
\label{fig:scattering}
\end{figure}

In Figure~\ref{fig:scattering}, we plot the bounds on spin-independent (left panel) and spin-dependent (right panel) cross sections mediated by operators which allow velocity independent scattering in addition to $\Upsilon (1S) \rightarrow \gamma + invisible$ decays.
We also plot 95\% CL bounds arising from Fermi-LAT searches for dark matter annihilation in dwarf spheroidal galaxies
and 90\% CL bounds arising from monojet and monophoton searches (ATLAS~\cite{Aad:2015zva,Aad:2014tda} and CMS~\cite{Khachatryan:2014rra, Khachatryan:2014rwa}).
The 90\% CL exclusion contours from CRESST II~\cite{Angloher:2015ewa},
SuperCDMS~\cite{CDMS:2014}, LUX~\cite{Akerib:2013tjd}, PICO~\cite{Amole:2015lsj} and PICASSO~\cite{Archambault:2012pm}
are also shown, as are the DAMA/LIBRA~\cite{Savage:2008er}, CRESST II (95\% CL)~\cite{Angloher:2011uu}, CoGeNT~\cite{Aalseth:2014jpa} and
CDMS II(Silicon)~\cite{Agnese:2013rvf} 90\% CL signal regions.

\section{Conclusions}

We have calculated the branching fractions for bound state quarkonium decays to final states with two dark matter particles and a photon. Noting the previous results for purely invisible final states~\cite{Fernandez:2014eja},  $\Upsilon (1S)$ decays can constrain all possible dimension 6 or lower effective contact operators coupling quarks to light dark matter of spin-0, spin-1/2 or spin-1. In particular, the possible complementarity of constraints between the decays of various  bound state mesons, along with a choice of final state, offers a unique handle on the characteristics of light dark matter interacting with heavy flavor quarks. We have extrapolated the limits on the total branching fraction for $\Upsilon (1S) \rightarrow \gamma \bar X X$ decays for all relevant operators by approximating the limits on the partial branching fraction into the phase space of the detector considered in~\cite{delAmoSanchez:2010ac} and rescaling to limits on the full branching fraction, given the respective efficiencies of the analysis cuts, for each operator. Once we approximate limits on the full branching for $\Upsilon (1S) \rightarrow \gamma \bar X X$ decays, we then calculate the bounds on the respective suppression scale, $\Lambda$, for each operator.   

In addition to the complementary constraints between various bound state decays, we have also related constraints from meson decays to analogous monojet/monophoton searches at hadron colliders and to searches for photons from dark matter annihilation in dwarf spheroidal galaxies. If one assumes scalar and pseudoscalar couplings respect minimal flavor violation, constraints from heavy quarkonium decays can be nearly as stringent as other search strategies which probe interactions with light flavor quarks and, thus, couple to light quark masses. Collider constraints, in general, are not threshold limited and can probe a dark matter mass range which can prove challenging to access when observing nuclear recoils at direct detection experiments or Standard Model decay products at indirect dark matter searches. Also, the contact operator approximation breaks down as the mediation scale approaches the characteristic energy scale of the process. While LHC searches can still set stringent limits on dark matter-Standard Model interactions through mediators with masses $\lsim {\cal O} ( \tev )$, bound state decays allow for model independent constraints to be set for mediator masses $\gsim 10 \gev$. Together with possibility of discriminating between effective operators based on different initial/final combinations, heavy quarkonium decays can augment model independent searches for dark matter at LHC.   

Future analysis of heavy quarkonium decays at high luminosity $e^+ / e^-$ colliders should not only improve sensitivity to $invisible$ and $invisble + \gamma$ final states, but also yield limits for decays of bound states beyond the more common $J^{P C} = 1^{- -}$   
mesons. Improved bounds from an upcoming Belle II analysis should enhance sensitivity to $\Upsilon (1S) \rightarrow \gamma \bar X X$ decays by a factor $\sim 4$~\cite{Browder}. Also, with enough data, Belle II will be capable of setting the first limits on $ h_b (1P) \rightarrow \gamma \bar X X$  decays, which, as we have shown, can be used to constrain particular subsets of effective operators allowing for $\Upsilon (1S) \rightarrow \gamma \bar X X$ decays. Heavy quarkonium decays at high luminosity experiments will continue to provide interesting constraints on Standard Model interactions with low mass dark matter candidates.

{\bf Acknowledgements}

We are grateful to J.~Kumar, S.~Profumo, X.~Tata, and S.~Vahsen for useful discussions.
This work is supported in part by Department of Energy grant  DE-SC0010504.

\appendix
\section{Constraints from Operators Mediating Purely Invisible Decays}

In a similar framework, the authors of~\cite{Fernandez:2014eja} have studied $\Upsilon (1S)$ (and $J / \Psi$) decays to $invisible$, emphasizing the complementarity of dark matter constraints from direct, indirect and collider searches. We update the results here to demonstrate the relationship between the decays of different bound states to either $invisible$ or $\gamma + invisible$ final states. We also calculate the invisible branching fractions and annihilation cross section relevant for the V7$_-$ and V9$_-$ operators which were not considered in previous work. Assuming weak interactions are negligible, the quark bilinears in the operators mediating decays to  a purely invisible final state must share the angular momentum and $C/P$ properties of the heavy quarkonium state. Specifically, for $\Upsilon (1S) $ mesons with $J^{PC} = 1^{--}$, the quark bilinear must be $\bar q \gamma^i q$ or $\bar q \sigma^{0i} q$, where $i$ is a spatial index~\cite{Kumar:2013iva}. Thus, the effective operators which can yield a nonvanishing matrix element for $\Upsilon (1S) \rightarrow \bar X X$ decays are orthogonal to the operators which allow $\Upsilon (1S) \rightarrow \gamma \bar X X$ decays.
The interaction structures permitting $J^{PC} = 1^{--}$ bound state decays to $invisible$ are listed in Table~\ref{tab:structsINV}, along with the angular momentum and $C/P$ properties of other possible bound states with non-vanishing matrix elements for decays to $\rightarrow  invisible$ or $\rightarrow \gamma + invisible$ final states. If the interaction structures can permit $s$-wave dark matter annihilation, then a bound can also be set by Fermi observations of photons originating from dwarf spheroidal galaxies. Lastly, we indicate whether or not constraints on spin-independent (SI) or spin-dependent (SD) nucleon scattering can be inferred from BaBar or Fermi limits. 

\begin{table}[h]
\centering
\begin{tabular}{|c|c|c|c|c|c|}
\hline
 Name & Interaction Structure & Invisible & Radiative & Annihilation  & Scattering \\
\hline
 F5  & $ (1/\Lambda^2) \bar X \gamma^\mu X \bar q \gamma_\mu q$ & $(1^{- -} , 1 , 0)$ & $(0^{+ +} , 1 , 1) $ & Yes & SI \\
     & & & $(0^{- +} , 0 , 0) $ &  &    \\
         & &  & $ (1^{+ +} , 1 , 1) $  &  &  \\
\hline
 F6  & $ (1/\Lambda^2) \bar X \gamma^\mu  \gamma^5 X \bar q \gamma_\mu q$ & $(1^{- -} , 1 , 0)$  & $(0^{+ +} , 1 , 1) $ & No & No \\
     & & & $(0^{- +} , 0 , 0) $  &   &  \\
         & &  & $ (1^{+ +} , 1 , 1) $  &  &  \\
\hline
 F9  & $ (1/\Lambda^2) \bar X \sigma^{\mu \nu}  X \bar q \sigma_{\mu \nu}   q$ & $(1^{- -} , 1 , 0)$  & $(0^{+ +} , 1 , 1) $& Yes & SD \\
 & & $ (1^{+ -} , 0 , 1) $ & $(0^{- +} , 0 , 0) $ &  &  \\
\hline
 F10  & $ (1/\Lambda^2)  \bar X \sigma^{\mu \nu}   \gamma^5 X \bar q \sigma_{\mu \nu} q$ & $(1^{- -} , 1 , 0)$  & $(0^{+ +} , 1 , 1) $ & Yes & No \\
 & & $ (1^{+ -} , 0 , 1) $ & $(0^{- +} , 0 , 0) $  &  &  \\
\hline
 S3  & $ (1/ 2 \Lambda^2) \imath ( \phi^{\dagger} \partial_\mu \phi -  \phi \partial_\mu \phi^{\dagger} ) \bar q \gamma^\mu q $& $(1^{- -} , 1 , 0)$  & $(0^{+ +} , 1 , 1) $& No & SI \\
  & & & $(0^{- +} , 0 , 0) $  &  &  \\
      & &  & $ (1^{+ +} , 1 , 1) $  &  &  \\
\hline
 V3  & $ (1/ 2 \Lambda^2) \imath ( B^{\dagger}_\nu \partial_\mu B^\nu - B^\nu \partial_\mu B_\nu^{\dagger} ) \bar q \gamma^\mu q$ & $(1^{- -} , 1 , 0)$  & $(0^{+ +} , 1 , 1) $& No & SI \\
  & &  & $(0^{- +} , 0 , 0) $  &  &  \\ 
    & &  & $ (1^{+ +} , 1 , 1) $  &  &  \\
\hline
 V5  & $ (1 /  \Lambda) \imath B^{\dagger}_\mu B_\nu  \bar q \sigma^{\mu \nu} q$ & $(1^{- -} , 1 , 0)$  &$(0^{+ +} , 1 , 1) $& Yes & SD \\
 & & $ (1^{+ -} , 0 , 1) $ & $(0^{- +} , 0 , 0) $  &  &  \\
\hline
 V6  & $ (1/  \Lambda)  B^{\dagger}_\mu B_\nu \bar q \sigma^{\mu \nu} \gamma^5 q$ & $(1^{- -} , 1 , 0)$  & $(0^{+ +} , 1 , 1) $ & Yes & No \\
 & & $ (1^{+ -} , 0, 1) $ & $(0^{- +} , 0 , 0) $ &  &  \\
\hline
 V7$_+$  & $(1/ 2 \Lambda^2) ( B^{\dagger}_\nu \partial^\nu B_\mu + B_\nu \partial^\nu B_\mu^{\dagger} )  \bar q \gamma^\mu q $ &$(1^{- -} , 1 , 0)$   & $(0^{+ +} , 1 , 1) $ & No & No \\ 
 & &   & $(0^{- +} , 0 , 0) $  &  &  \\
     & &  & $ (1^{+ +} , 1 , 1) $  &  &  \\
     \hline
 V7$_-$  & $(1/ 2 \Lambda^2) \imath ( B^{\dagger}_\nu \partial^\nu B_\mu - B_\nu \partial^\nu B_\mu^{\dagger} )  \bar q \gamma^\mu q $ &$(1^{- -} , 1 , 0)$   & $(0^{+ +} , 1 , 1) $ & No & No \\ 
 & &   & $(0^{- +} , 0 , 0) $  &  &  \\
     & &  & $ (1^{+ +} , 1 , 1) $  &  &  \\
\hline
 V9$_+$   & $  (1/ 2 \Lambda^2)  \epsilon^{\mu \nu \rho \sigma} ( B^{\dagger}_\nu \partial_\rho B_\sigma + B_\nu \partial_\rho B_\sigma^{\dagger} ) \bar q \gamma_\mu  q$ & $(1^{- -} , 1 , 0)$   & $(0^{+ +} , 1 , 1) $& No & No \\
 & & & $(0^{- +} , 0 , 0) $ &  &  \\
     & &  & $ (1^{+ +} , 1 , 1) $  &  &  \\
     \hline
 V9$_-$   & $  (1/ 2 \Lambda^2) \imath \epsilon^{\mu \nu \rho \sigma} ( B^{\dagger}_\nu \partial_\rho B_\sigma - B_\nu \partial_\rho B_\sigma^{\dagger} ) \bar q \gamma_\mu  q$ & $(1^{- -} , 1 , 0)$   & $(0^{+ +} , 1 , 1) $& Yes & No \\
 & & & $(0^{- +} , 0 , 0) $ &  &  \\
     & &  & $ (1^{+ +} , 1 , 1) $  &  &  \\
\hline
\end{tabular}
\caption{Interaction structures that can mediate $\Upsilon (1S)$ decays to $invisible$. Invisible (radiative) identifies the $(J^{PC}, S, L)$ bound states that can be annihilated for $invisible$ ($invisible + \gamma$) final states. Note that we only consider $s$- and $p$-wave bound states. If the interaction structures can permit $s$-wave dark matter annihilation, then a bound can also be set by Fermi observations of photons originating from dwarf spheroidal galaxies. Lastly, we indicate whether or not constraints on spin-independent (SI) or spin-dependent (SD) nucleon scattering can be inferred from BaBar or Fermi limits.}
\label{tab:structsINV}
\end{table}

As noted in the text, the various DM-SM interaction structures will have complementarity constraints due to decays of different bound states to invisible final states. For example, assuming fermionic dark matter interacts through our F5 operator, which arises from integrating out a vector mediator, constraints from $\Upsilon (1S) \rightarrow  \bar X X$ decays will be directly related $\eta_b (1P) \rightarrow \gamma \bar X X$, $\chi_{b0} (1P) \rightarrow \gamma \bar X X$ and $\chi_{b1} (1P) \rightarrow \gamma \bar X X$ decays. As a result, between decays to $invisible$ and $invisible + \gamma$ final states, every low-lying bottomonium state can be used to constrain a variety of dark matter interaction structures.     

As pointed out by the authors of~\cite{Kumar:2015wya}, a complete set of dimension 6 or lower contact operators mediating spin-1 dark matter interactions with quarks should include $C$-odd V(7-10)$_-$ interaction structures in addition to the $C$-even V(7-10)$_+$ operators considered in~\cite{Kumar:2013iva}. We have included both groups of operators in our analysis of bound state decays to $\gamma +invisible$ in this work. We calculate the branching fractions for bound state decays to $invisible$ through the V7$_-$ and V9$_-$ operators not considered in~\cite{Fernandez:2014eja},    
\bea
{\cal B}_{V7_-}( \bar X X ) &=& 
{  {\cal B} ( e^+ e^- ) M^4  \over 512 \pi^2 \alpha^2 Q_b^2 \Lambda^4}
\left( 1 - {4 m_X^2 \over M^2} \right)^{3/2} { M^2 \over  m_X^2 } 
\left( 1 + { M^2 \over 4 m_X^2} \right) ,
\nonumber\\
{\cal B}_{V9_-}( \bar X X ) &=& 
{  {\cal B} ( e^+ e^- ) M^4  \over 128 \pi^2 \alpha^2 Q_b^2 \Lambda^4} 
\left( 1 - {4 m_X^2 \over M^2} \right)^{1/2}
\left( 1 + { M^2 \over 2 m_X^2} \right) .
\eea 
Note that, while the V(7-10)$_+$ operators allow for the dark matter fields to be real or complex, the V(7-10)$_-$ operators are only non-vanishing if we assume complex dark matter. The V7$_+$ and V9$_+$ branching fractions to $invisible$ calculated in~\cite{Fernandez:2014eja} also assumed operators with complex dark matter, albeit with an extra factor of 2 in the normalization of the operators. For completeness, we also calculate the dark matter annihilation cross section for the V9$_-$ interaction structure,  
\bea
\langle \sigma_A^{V9_-} v \rangle &=& { m_X^2 \over 3 \pi \Lambda^4 }  \left( 1 - { m_q^2 \over m_X^2} \right)^{1/2} \left( 1 + { m_q^2 \over 2 m_X^2 } \right) .
\eea

\section{Polarized Differential Branching Fractions}

We calculate the differential branching fractions for $\Upsilon (1S) \rightarrow \gamma \bar X X$, with respect to photon energy and scattering angle, assuming polarized $\Upsilon (1S)$ produced in $\Upsilon (2S) \rightarrow \pi^+ \pi^- \Upsilon (1S)$ transitions. For relativistic $e^+ e^-$ annihilating through a photon, the resulting $\Upsilon (2S)$ is produced polarized with its spin axis lying along the beam line. We assume the dipion transition is dominated by $E1 \cdot E1$ gluon radiation, thus the daughter $\Upsilon (1S)$ should also be polarized along the beam line. We also assume a negligible differential boost between the rest frame of the $\Upsilon (2S)$ and that of the $\Upsilon (1S)$. The polarization of the $\Upsilon (1S)$ and the lack of relative angular momentum between the $\Upsilon (1S)$ and the dipion system have been confirmed in the angular distributions of charged leptons produced in subsequent $\Upsilon (1S)$ decays and in the dipion system, respectively~\cite{Alexander:1998dq}.

We can define the kinematics of the $\Upsilon (1S) \rightarrow \gamma X$ decay, where $X$ now denotes the invisible system consisting of our dark matter particles. Since we are considering the rest frame of the $\Upsilon (2S)$, and, to a reasonable approximation, that of the $\Upsilon (1S)$, we define $P_\Upsilon = (M, 0 , 0, 0)$. The photon will be emitted at an angle, $\theta$, relative to the beam line with an energy, $\omega$, yielding $k_\gamma = (\omega, \omega \sin \theta, 0 , \omega \cos \theta )$. Finally, we only consider $\Upsilon (1S)$ polarizations with spin projections on the beam axis, $\epsilon_\Upsilon^{\pm} = \mp ~ 2^{- 1/ 2} (0, 1, \pm \imath, 0) $. Now we can calculate  the matrix elements squared for our polarized $\Upsilon (1S)$ decays, given the relevant DM-SM interaction structures, and then integrate over the full phase space of the invisible system, leaving phase space distributions only in our observable kinematic variables.  

We write the differential branching fractions using abbreviated notation for some of the phase space factors,
\bea
\lambda (a, b, c) &=& a^2 + b^2 + c^2 - 2 a b - 2 ac - 2 b c ,  
\eea
with $\lambda_\Upsilon = \lambda (1, X^2 / M^2, 0)$ and $\lambda_X = \lambda (1, m_X^2 / X^2, m_X^2 / X^2)$. We also denote the possible angular dependencies for  $\Upsilon (1S) \rightarrow \gamma \bar X X$ decays in our framework, $f_\theta^+ = 1 + \cos^2 \theta $ and $f_\theta^- = 2 - 2 \cos^2 \theta$. The following are the polarized differential branching fractions as functions of dark matter mass, $\Upsilon (1S)$ mass, mediation scale, the branching fraction to $e^+ / e^-$, and phase space factors:   
\bea
 { d {\cal B}_{F1,F3}^{pol} (\gamma \bar X X ) \over d X^2 d \cos \theta } &=&
  { 3 {\cal B} ( e^+ e^- ) M^2  \over 512 \pi^3 \alpha \Lambda^6}
\lambda_\Upsilon^{1/2} \lambda_X^{3/2}  X^2 
f_\theta^+ ,
 \nonumber\\
 { d {\cal B}_{F2,F4}^{pol} (\gamma \bar X X ) \over d X^2 d \cos \theta } &=&
  { 3 {\cal B} ( e^+ e^- ) M^2  \over 512 \pi^3 \alpha \Lambda^6}
\lambda_\Upsilon^{1/2} \lambda_X^{1/2}   X^2 
f_\theta^+ ,
  \nonumber\\
  { d {\cal B}_{F7}^{pol} (\gamma \bar X X ) \over d X^2 d \cos \theta } &=&
  { 3 {\cal B} ( e^+ e^- )   \over 128 \pi^3 \alpha \Lambda^4}
  \lambda_\Upsilon^{1/2}
  \left[\left(   - {X^2 \over 3} \lambda_X^{3/2} + X^2 \lambda_X^{1/2} \right) 
  f_\theta^-
 +  {2 \over 3} M^2 \left( 1 + 2 {m_X^2 \over X^2} \right) 
  \lambda_X^{1/2}
  f_\theta^+
  \right] ,
  \nonumber\\
 { d {\cal B}_{F8}^{pol} (\gamma \bar X X ) \over d X^2 d \cos \theta } &=&
  {  {\cal B} ( e^+ e^- )  \over 64 \pi^3 \alpha \Lambda^4}
  \lambda_\Upsilon^{1/2}
  \left[    X^2 \lambda_X^{3/2}  
  f_\theta^-
 +  M^2 \left( 1 + 2 {m_X^2 \over X^2} \right) 
  \lambda_X^{1/2}
  f_\theta^+
  \right] ,
  \nonumber\\
  { d {\cal B}_{S1,S2}^{pol} (\gamma \bar X X ) \over d X^2 d \cos \theta } &=&
  { 3 {\cal B} ( e^+ e^- ) M^2  \over 1024 \pi^3 \alpha \Lambda^4}
\lambda_\Upsilon^{1/2} \lambda_X^{1/2}
f_\theta^+ ,
 \nonumber \\
  { d {\cal B}_{S4}^{pol} (\gamma \bar X X ) \over d X^2 d \cos \theta } &=&
  {  {\cal B} ( e^+ e^- ) M^2  \over 1024 \pi^3 \alpha \Lambda^4}
\lambda_\Upsilon^{1/2}
\lambda_X^{3/2}
\left[  {X^2 \over M^2} f_\theta^- +   f_\theta^+   \right]  ,
\nonumber\\
  { d {\cal B}_{V1,V2}^{pol} (\gamma \bar X X ) \over d X^2 d \cos \theta } &=&
  { 3 {\cal B} ( e^+ e^- ) M^2  \over 1024 \pi^3 \alpha \Lambda^4}
\lambda_\Upsilon^{1/2} 
 \lambda_X^{1/2}
 \left[ 3 - {X^2 \over m_X^2} + {X^4 \over 4 m_X^4} \right]
   f_\theta^+ ,
   \nonumber\\
   { d {\cal B}_{V4}^{pol} (\gamma \bar X X ) \over d X^2 d \cos \theta } &=&
  {  {\cal B} ( e^+ e^- ) M^2  \over 1024 \pi^3 \alpha \Lambda^4}
\lambda_\Upsilon^{1/2}  \lambda_X^{3/2}
\left[ 3 - {X^2 \over m_X^2} + {X^4 \over 4 m_X^4} \right] 
\left[  {X^2 \over M^2} f_\theta^-  +   f_\theta^+    \right] ,
\nonumber\\
   { d {\cal B}_{V8_+}^{pol} (\gamma \bar X X ) \over d X^2 d \cos \theta } &=&
 {  {\cal B} ( e^+ e^- )   \over 4096 \pi^3 \alpha \Lambda^4}
  \lambda_\Upsilon^{1/2} \lambda_X^{3/2} 
  \left[   4 {X^4 \over m_X^2}  
  f_\theta^- 
 +  M^2 \left( 3 {X^4 \over m_X^4} - 8 {X^2 \over m_X^2}  \right)  
  f_\theta^+
  \right] ,
  \nonumber\\
  { d {\cal B}_{V8_-}^{pol} (\gamma \bar X X ) \over d X^2 d \cos \theta } &=& 
 {  {\cal B} ( e^+ e^- ) M^2   \over 1024 \pi^3 \alpha \Lambda^4}
 \lambda_\Upsilon^{1/2}  \lambda_X^{3/2} 
 {X^2 \over m_X^2}  \left[   1+ {X^2 \over 4 m_X^2}  
\right]
\left[  {X^2 \over M^2} f_\theta^-  +   f_\theta^+    \right] ,
  \nonumber\\
   { d {\cal B}_{V10_+}^{pol} (\gamma \bar X X ) \over d X^2 d \cos \theta } &=&
  {  {\cal B} ( e^+ e^- )   \over 1024 \pi^3 \alpha \Lambda^4}
 \lambda_\Upsilon^{1/2} 
  \left[    {X^4 \over m_X^2} \lambda_X^{5/2}  
 f_\theta^- 
 +  M^2 {X^2 \over m_X^2}  \left( 1 + 2 {m_X^2 \over X^2} \right) 
  \lambda_X^{3/2}
 f_\theta^+
  \right] ,
  \nonumber\\
   { d {\cal B}_{V10_-}^{pol} (\gamma \bar X X ) \over d X^2 d \cos \theta } &=&  {  {\cal B} ( e^+ e^- ) M^2   \over 512 \pi^3 \alpha \Lambda^4}
  \lambda_\Upsilon^{1/2}  \lambda_X^{1/2} 
  \left[   1+ {X^2 \over 2 m_X^2}  
\right]
\left[  {X^2 \over M^2} f_\theta^-  +   f_\theta^+    \right] .
\eea
Note if we integrate these differential branching fractions over the entire kinematic range, which is equivalent to assuming an isotropically polarized bound state, we recover the fully integrated branching fractions reported in eq.~\ref{eqn:fullwidths}.

\section{Phase Space Integrals} \label{phase_space}

We show the analytic results of the relevant integrals over the phase space of the invariant mass squared of the invisible system, $X^2$, noting the change of integration variables to $x ' = X^2 / 4 m_X^2$. The integrals are all of the form
\bea
I_n^m (x) &=& \int_{1}^{x} \left( 1 - {1 \over x '} \right)^n x'^{~m} dx', 
\eea
where $n=1/2, 3/2$ and $m$ is an integer such that $-1 \le m \le 4 $.
For calculation of the partially integrated branching fractions, the allowed range of $X^2$ is determined by the photon detection threshold, $\omega_{min}$, as well as the dark matter mass, such that $4 m_X^2 < X^2 < M^2 - 2 M \omega_{min}$. The full branching fractions are integrated over the entire kinematic range with $\omega_{min} = 0$.       
\bea
I_{1/2}^{-1} (x)
&=& - 2
\left[
\sqrt{1 - { 1 \over x} }
+  \log \left( \sqrt{x -1} - \sqrt{x} \right)
\right] 
\nonumber\\
I_{1/2}^0 (x)
&=& 
\left[
\sqrt{x -1} \sqrt{x} 
- \log \left( \sqrt{x -1} + \sqrt{x} \right)
\right] 
\nonumber\\
I_{1/2}^1 (x)
&=& {1 \over 4} 
\left[
\sqrt{x -1} \sqrt{x} \left( 2 x - 1 \right)
- \log \left( \sqrt{x -1} + \sqrt{x} \right)
\right] 
\nonumber\\
I_{1/2}^2 (x)
&=& {1 \over 24} 
\left[
\sqrt{x -1} \sqrt{x} \left( 8 x^2 - 2 x - 3 \right)
- 3  \log \left( \sqrt{x -1} + \sqrt{x} \right)
\right] 
\nonumber\\
I_{1/2}^3 (x)
&=& {1 \over 192} 
\left[
\sqrt{x -1} \sqrt{x} \left(48 x^3 - 8 x^2 - 10 x - 15 \right)
- 15  \log \left( \sqrt{x -1} + \sqrt{x} \right)
\right] 
\nonumber
\nonumber\\
I_{1/2}^4 (x)
&=& {1 \over 1920} 
\left[
\sqrt{x -1} \sqrt{x} \left( 384 x^4 - 48 x^3 - 56 x^2 - 70 x - 105 \right)
- 105  \log \left( \sqrt{x -1} + \sqrt{x} \right)
\right] 
\nonumber\\
I_{3/2}^{-1} (x)
&=&  2
\left[
{ \sqrt{x -1 } \left(1 - 4 x \right) \over 3 x^{3/2} }
+  \log \left( \sqrt{x -1} + \sqrt{x} \right)
\right] 
\nonumber\\
I_{3/2}^0 (x)
&=& 
\left[
\sqrt{1 - { 1 \over x} } \left( x+ 2 \right) 
- 3 \log \left( \sqrt{x -1} + \sqrt{x} \right)
\right] 
\nonumber\\
I_{3/2}^1 (x)
&=& {1 \over 4} 
\left[
\sqrt{x -1} \sqrt{x} \left( 2 x - 5 \right)
+ 3 \log \left( \sqrt{x -1} + \sqrt{x} \right)
\right] 
\nonumber\\
I_{3/2}^2 (x)
&=& {1 \over 24} 
\left[
\sqrt{x -1} \sqrt{x} \left( 8 x^2 - 14 x + 3 \right)
+ 3  \log \left( \sqrt{x -1} + \sqrt{x} \right)
\right] 
\nonumber\\
I_{3/2}^3 (x)
&=& {1 \over 64} 
\left[
\sqrt{x -1} \sqrt{x} \left( 16 x^3 - 24 x^2 + 2 x + 3 \right)
+ 3  \log \left( \sqrt{x -1} + \sqrt{x} \right)
\right] 
\nonumber\\
I_{3/2}^4 (x)
&=& {1 \over 640} 
\left[
\sqrt{x -1} \sqrt{x} \left(128 x^4 - 176 x^3 + 8 x^2 + 10 x + 15 \right)
+ 15  \log \left( \sqrt{x -1} + \sqrt{x} \right)
\right] 
\eea
Note that $I_n^m (1) = 0$ for all $n$ and $m$, thus the integrals need only be evaluated at $x = ( M^2 - 2 M \omega_{min} ) / 4 m_X^2$.


\begin{thebibliography}{99}
\bibitem{Kolb:1990}
 E.~Kolb and M.~Turner,
   Front. Phys. {\bf 69}, 1-547 (1990).

\bibitem{Hooper:2009}
  M.~Beltran, D.~Hooper, E.~W.~Kolb and Z.~C.~Krusberg,
  Phys.\ Rev.\ D {\bf 80}, 043509 (2009)
  [arXiv:0808.3384 [hep-ph]].

\bibitem{Cao:2011}
  Q.~-H.~Cao, C.~-R.~Chen, C.~S.~Li and H.~Zhang,
  JHEP {\bf 1108}, 018 (2011)
  [arXiv:0912.4511 [hep-ph]].

\bibitem{Goodman:2011}
  J.~Goodman, M.~Ibe, A.~Rajaraman, W.~Shepherd, T.~Tait and H.~Yu,
  Phys.\ Lett.\ B {\bf 695}, 185 (2011)
  [arXiv:1005.1286 [hep-ph]].

\bibitem{Rajaraman:2013}
  A.~Rajaraman, T.~M.~P.~Tait and A.~M.~Wijangco,
  Phys.\ Dark Univ.\ {\bf 2}, 17 (2013)
  [arXiv:1211.7061 [hep-ph]].

\bibitem{Dreiner:2013}
  H.~Dreiner, D.~Schmeier and J.~Tattersall,
  Europhys.\ Lett.\ {\bf 102}, 51001 (2013)
  [arXiv:1303.3348 [hep-ph]].

\bibitem{Kumar:2013iva}
  J.~Kumar and D.~Marfatia,
  Phys.\  Rev.\ D {\bf 88}, 014035 (2013)
  [arXiv:1305.1611 [hep-ph]].

\bibitem{Busoni:2014}
  G.~Busoni, A.~De~Simone, E.~Morgante, and A.~Riotto,
  Phys.\ Lett.\ B {\bf 728}, 412 (2014)
  [arXiv:1307.2253 [hep-ph]].

\bibitem{Rajaraman:2013-2}
  A.~DiFranzo, K.~I.~Nagao, A.~Rajaraman and T.~M.~P.~Tait,
  JHEP {\bf 1311}, 014 (2013)
  [arXiv:1308.2679 [hep-ph]].

\bibitem{Buckley:2013}
  M.~R.~Buckley,
  Phys.\ Rev.\ D {\bf 88}, 055028 (2013)
  [arXiv:1308.4146 [hep-ph]].

\bibitem{Buchmueller:2014}
  O.~Buchmueller, M.~J.~Dolan, and C.~McCabe,
  JHEP {\bf 1401}, 025 (2014)
  [arXiv:1308.6799 [hep-ph]].

\bibitem{Busoni:2014-2}
  G.~Busoni, A.~De~Simone, J.~Gramling, E.~Morgante, and A.~Riotto,
  [arXiv:1402.1275 [hep-ph]].

\bibitem{Alves:2014}
   A.~Alves, S.~Profumo, F.~S.~Queiroz, and W.~Shepherd,
  [arXiv:1403.5027 [hep-ph]].

\bibitem{Fedderke:2014}
   M.~A.~Fedderke, J.-Y.~Chen, E.~W.~Kolb, and L.-T.~Wang,
  [arXiv:1404.2283 [hep-ph]].

\bibitem{D'Eramo:2014aba} 
  F.~D'Eramo and M.~Procura,
  JHEP {\bf 1504}, 054 (2015)
  doi:10.1007/JHEP04(2015)054
  [arXiv:1411.3342 [hep-ph]].


 \bibitem{Birkedal:2004}
   A.~Birkedal, K.~Matchev and M.~Perelstein,
   Phys.\ Rev.\ D {\bf 70}, 077701 (2004)
   [arXiv:hep-ph/0403004].

\bibitem{Feng:2005gj}
  J.~L.~Feng, S.~Su and F.~Takayama,
  Phys.\ Rev.\ Lett.\  {\bf 96}, 151802 (2006)
  [hep-ph/0503117].


 \bibitem{Hooper:2010}
    M.~Beltran, D.~Hooper, E.~W.~Kolb, Z.~C.~Krusberg and T.~M.~P.~Tait,
   JHEP {\bf 1009}, 037 (2010)
   [arXiv:1002.4137 [hep-ph]].

 \bibitem{Bai:2010}
    Y.~Bai, P.~J.~Fox and R.~Harnik,
   JHEP {\bf 1012}, 048 (2010)
   [arXiv:1005.3797 [hep-ph]].

\bibitem{Goodman:2010}
  J.~Goodman, M.~Ibe, A.~Rajaraman, W.~Shepherd, T.~Tait and H.~Yu,
  Phys.\  Rev.\ D {\bf 82}, 116010 (2010)
  [arXiv:1008.1783 [hep-ph]].

 \bibitem{Rajaraman:2011}
   A.~Rajaraman, W.~Shepherd, T.~M.~P.~Tait and A.~M.~Wijangco,
   Phys.\ Rev.\ D {\bf 84}, 095013 (2011)
   [arXiv:1108.1196 [hep-ph]].

 \bibitem{Fox:2012}
   P.~J.~Fox, R.~Harnik, J.~Kopp and Y.~Tsai,
   Phys.\ Rev.\ D {\bf 85}, 056011 (2012)
   [arXiv:1109.4398 [hep-ph]].

 \bibitem{Bai:2011}
   Y.~Bai, A.~Rajaraman,
   [arXiv: 1109.6009 [hep-ph]].

 \bibitem{Goodman:2011-2}
   J.~Goodman and W.~Shepherd,
   [arXiv:1111.2359 [hep-ph]].

 \bibitem{Bai:2013}
   Y.~Bai and T.~M.~Tait,
   Phys.\ Lett.\ B {\bf 723}, 384 (2013)
   [arXiv:1208.4361 [hep-ph]].

 \bibitem{Agrawal:2013}
   P.~Agrawal and V.~Rentala,
   [arXiv:1312.5325 [hep-ph]].

 \bibitem{Papucci:2014}
   M.~Papucci, A.~Vichi, and K.~M.~Zurek,
   [arXiv: 1402.2285 [hep-ph]].

\bibitem{Kumar:2015wya} 
  J.~Kumar, D.~Marfatia and D.~Yaylali,
  arXiv:1508.04466 [hep-ph].


\bibitem{Essig:2013}
R.~Essig, J.~Mardon, M.~Papucci, T.~Volansky and Y.~Zhong,
  [arXiv:1309.5084 [hep-ph]].

\bibitem{Fayet:2007}
  P.~Fayet,
  Phys.\  Rev.\ D {\bf 75}, 115017 (2007)
  [hep-ph/0702176].

\bibitem{Fayet:2010}
  P.~Fayet,
  Phys.\  Rev.\ D {\bf 81}, 054025 (2010)
  [arXiv:0910.2587 [hep-ph]].

\bibitem{McElrath:2007}
  B.~McElrath,
  [arXiv:0712.0016 [hep-ph]].

\bibitem{Cotta:2013}
  R.~Cotta, A.~Rajaraman, T.~Tait and A.~Wijangco,
  [arXiv:1305.6609 [hep-ph]].

\bibitem{Schmidt-Hoberg:2013hba}
  K.~Schmidt-Hoberg, F.~Staub and M.~W.~Winkler,
  Phys.\ Lett.\ B {\bf 727}, 506 (2013)
  [arXiv:1310.6752 [hep-ph]].

\bibitem{Badin:2010}
  A.~Badin and A.~Petrov,
  [arXiv:1005.1277 [hep-ph]].

\bibitem{McKeen:2009rm}
  D.~McKeen,
  Phys.\ Rev.\ D {\bf 79}, 114001 (2009)
  [arXiv:0903.4982 [hep-ph]].

\bibitem{Fernandez:2014eja} 
  N.~Fernandez, J.~Kumar, I.~Seong and P.~Stengel,
  Phys.\ Rev.\ D {\bf 90}, no. 1, 015029 (2014)
  [arXiv:1404.6599 [hep-ph]].

\bibitem{Yeghiyan:2010}
  G.~Yeghiyan,
  Phys.\  Rev.\ D {\bf 80}, 115019 (2009)
  [arXiv:0909.4919 [hep-ph]].

\bibitem{Aubert:2008as} 
  B.~Aubert {\it et al.} [BaBar Collaboration],
  [arXiv:0808.0017 [hep-ex]].

\bibitem{PDG}
J.~Beringer {\it et al.} (Particle Data Group)
 Phys.\ Rev.\ D {\bf 86}, 010001 (2012).

\bibitem{Balest:1994ch} 
  R.~Balest {\it et al.} [CLEO Collaboration],
  Phys.\ Rev.\ D {\bf 51}, 2053 (1995).

\bibitem{delAmoSanchez:2010ac} 
  P.~del Amo Sanchez {\it et al.} [BaBar Collaboration],
  Phys.\ Rev.\ Lett.\  {\bf 107}, 021804 (2011)
  [arXiv:1007.4646 [hep-ex]].

\bibitem{Essig:2009}
  R.~Essig, N.~Sehgal and L.~Strigari,
     Phys.\  Rev.\  D {\bf 80}, 023506 (2009)
  [arXiv:0902.4750 [hep-ph]].

\bibitem{Geringer:2011}
  A.~Geringer-Sameth and S.~Koushiappas,
   Phys.\  Rev.\ Lett. {\bf 107}, 241303 (2011)
  [arXiv:1108.2914 [astro-ph.CO]].

\bibitem{Geringer:2012}
  A.~Geringer-Sameth and S.~Koushiappas,
   Phys.\  Rev.\  D {\bf 86}, 021302(R) (2012)
  [arXiv:1206.0796 [astro-ph.HE]].

\bibitem{Fermi:2011}
  The Fermi-LAT Collaboration,
   Phys.\  Rev.\ Lett. {\bf 107}, 241302 (2011)
  [arXiv:1108.3546[astro-ph.HE]].

\bibitem{Fermi:2013}
  The Fermi-LAT Collaboration,
  Phys.\  Rev.\ D {\bf 89}, 042001 (2014)
  [arXiv:1310.0828 [astro-ph.HE]].

\bibitem{Ackermann:2015zua} 
  M.~Ackermann {\it et al.} [Fermi-LAT Collaboration],
  Phys.\ Rev.\ Lett.\  {\bf 115}, no. 23, 231301 (2015)
  doi:10.1103/PhysRevLett.115.231301
  [arXiv:1503.02641 [astro-ph.HE]].

\bibitem{Baring:2015sza} 
  M.~G.~Baring, T.~Ghosh, F.~S.~Queiroz and K.~Sinha,
  arXiv:1510.00389 [hep-ph].

\bibitem{Pythia}
  T. Sjöstrand, S. Mrenna and P. Skands,
  JHEP {\bf 0605}, 026 (2006),
  [arXiv:hep-ph/0603175].

\bibitem{Gunion:2005rw} 
  J.~F.~Gunion, D.~Hooper and B.~McElrath,
  Phys.\ Rev.\ D {\bf 73}, 015011 (2006)
  [hep-ph/0509024].

\bibitem{Haisch:2013uaa}
  U.~Haisch and F.~Kahlhoefer,
  JCAP {\bf 1304}, 050 (2013)
  [arXiv:1302.4454 [hep-ph]].

\bibitem{Crivellin:2014qxa}
  A.~Crivellin, F.~D'Eramo and M.~Procura,
  Phys.\ Rev.\ Lett.\  {\bf 112}, 191304 (2014)
  [arXiv:1402.1173 [hep-ph]].

\bibitem{Aad:2015zva} 
  G.~Aad {\it et al.} [ATLAS Collaboration],
  Eur.\ Phys.\ J.\ C {\bf 75}, no. 7, 299 (2015)
  [Eur.\ Phys.\ J.\ C {\bf 75}, no. 9, 408 (2015)]
  [arXiv:1502.01518 [hep-ex]].

\bibitem{Aad:2014tda} 
  G.~Aad {\it et al.} [ATLAS Collaboration],
  Phys.\ Rev.\ D {\bf 91}, no. 1, 012008 (2015)
  [arXiv:1411.1559 [hep-ex]].

\bibitem{Khachatryan:2014rra} 
  V.~Khachatryan {\it et al.} [CMS Collaboration],
  Eur.\ Phys.\ J.\ C {\bf 75}, no. 5, 235 (2015)
  [arXiv:1408.3583 [hep-ex]].

\bibitem{Khachatryan:2014rwa} 
  V.~Khachatryan {\it et al.} [CMS Collaboration],
  arXiv:1410.8812 [hep-ex].

\bibitem{Haisch:2015ioa} 
  U.~Haisch and E.~Re,
  JHEP {\bf 1506}, 078 (2015)
  doi:10.1007/JHEP06(2015)078
  [arXiv:1503.00691 [hep-ph]].

\bibitem{Buchmueller:2014yoa} 
  O.~Buchmueller, M.~J.~Dolan, S.~A.~Malik and C.~McCabe,
  JHEP {\bf 1501}, 037 (2015)
  doi:10.1007/JHEP01(2015)037
  [arXiv:1407.8257 [hep-ph]].

\bibitem{Kelso:2014qja} 
  C.~Kelso, J.~Kumar, P.~Sandick and P.~Stengel,
  Phys.\ Rev.\ D {\bf 91}, 055028 (2015)
  [arXiv:1411.2634 [hep-ph]].

\bibitem{Alarcon:2011zs} 
  J.~M.~Alarcon, J.~Martin Camalich and J.~A.~Oller,
  Phys.\ Rev.\ D {\bf 85}, 051503 (2012)
  doi:10.1103/PhysRevD.85.051503
  [arXiv:1110.3797 [hep-ph]].

\bibitem{Alarcon:2012nr} 
  J.~M.~Alarcon, L.~S.~Geng, J.~Martin Camalich and J.~A.~Oller,
  Phys.\ Lett.\ B {\bf 730}, 342 (2014)
  doi:10.1016/j.physletb.2014.01.065
  [arXiv:1209.2870 [hep-ph]].

\bibitem{Hoferichter:2015dsa} 
  M.~Hoferichter, J.~Ruiz de Elvira, B.~Kubis and U.~G.~Meißner,
  Phys.\ Rev.\ Lett.\  {\bf 115}, no. 9, 092301 (2015)
  doi:10.1103/PhysRevLett.115.092301
  [arXiv:1506.04142 [hep-ph]].

\bibitem{Ellis:2009ai} 
  J.~Ellis, K.~A.~Olive and P.~Sandick,
  New J.\ Phys.\  {\bf 11}, 105015 (2009)
  [arXiv:0905.0107 [hep-ph]].

\bibitem{Angloher:2015ewa} 
  G.~Angloher {\it et al.} [CRESST Collaboration],
  arXiv:1509.01515 [astro-ph.CO].

\bibitem{CDMS:2014}
  R.~Agnese et al.,
  [arXiv:1402.7137 [hep-ex]].

\bibitem{Akerib:2013tjd}
  D.~S.~Akerib {\it et al.}  [LUX Collaboration],
  arXiv:1310.8214 [astro-ph.CO].

\bibitem{Amole:2015lsj} 
  C.~Amole {\it et al.} [PICO Collaboration],
  Phys.\ Rev.\ Lett.\  {\bf 114}, no. 23, 231302 (2015)
  [arXiv:1503.00008 [astro-ph.CO]].

\bibitem{Archambault:2012pm}
  S.~Archambault {\it et al.}  [PICASSO Collaboration],
  Phys.\ Lett.\ B {\bf 711}, 153 (2012)
  [arXiv:1202.1240 [hep-ex]].

\bibitem{Savage:2008er}
  C.~Savage, G.~Gelmini, P.~Gondolo and K.~Freese,
  JCAP {\bf 0904}, 010 (2009)
  [arXiv:0808.3607 [astro-ph]].

\bibitem{Angloher:2011uu}
  G.~Angloher, M.~Bauer, I.~Bavykina, A.~Bento, C.~Bucci, C.~Ciemniak, G.~Deuter and F.~von Feilitzsch {\it et al.},
  Eur.\ Phys.\ J.\ C {\bf 72}, 1971 (2012)
  [arXiv:1109.0702 [astro-ph.CO]].

\bibitem{Aalseth:2014jpa} 
  C.~E.~Aalseth {\it et al.},
  arXiv:1401.6234 [astro-ph.CO].

\bibitem{Agnese:2013rvf}
  R.~Agnese {\it et al.}  [CDMS Collaboration],
  Phys.\ Rev.\ Lett.\  {\bf 111}, 251301 (2013)
  [arXiv:1304.4279 [hep-ex]].

 

\bibitem{Browder}
T.~Browder, private communication.


\bibitem{Alexander:1998dq} 
  J.~P.~Alexander {\it et al.} [CLEO Collaboration],
  Phys.\ Rev.\ D {\bf 58}, 052004 (1998)
  [hep-ex/9802024].







\end{thebibliography}
\end{document}